\documentclass[10pt,twocolumn,preprintnumbers,amsmath,amssymb,nofootinbib
,superscriptaddress]{revtex4-1}
\usepackage{graphicx,longtable,mathrsfs,color,array}
\usepackage{hyperref}
\usepackage[usenames,dvipsnames]{xcolor} 
\usepackage{amssymb,amsmath,mathtools,mathrsfs,slashed} 
\usepackage{epsfig,subfigure,placeins,float} 
\usepackage{booktabs,longtable,ctable,multirow} 
\usepackage{exscale,relsize} 
\usepackage[normalem]{ulem} 
\usepackage{enumerate}
\usepackage{enumitem}
\usepackage{comment}
\usepackage{color}
\allowdisplaybreaks[1]

\newcommand{\REVIEWRELEASENUMBER}{LLNL-JRNL-857502-DRAFT}

\begin{document}
\title{Birefringence tests of gravity with multi-messenger binaries}
\author{Macarena Lagos}
\affiliation{Department of Physics and Astronomy, Columbia University, New York, NY 10027, USA}
\affiliation{Instituto de Astrof\'isica, Departamento de Ciencias F\'isicas, Universidad Andr\'es Bello, Santiago, Chile}
\author{Leah Jenks}
\affiliation{Kavli Institute for Cosmological Physics,
University of Chicago, Chicago, IL 60637, USA}
\author{Maximiliano Isi}
\affiliation{Center for Computational Astrophysics, Flatiron Institute,
162 5th Ave, New York, NY 10010,USA}
\author{Kenta Hotokezaka}
\affiliation{Research Center for the Early Universe, School of Science, The University of Tokyo, Bunkyo, Tokyo 113-0033, Japan}
\author{Brian D.~Metzger}
\affiliation{Department of Physics and Astronomy, Columbia University, New York, NY 10027, USA}
\affiliation{Center for Computational Astrophysics, Flatiron Institute,
162 5th Ave, New York, NY 10010,USA}
\author{Eric Burns}
\affiliation{Department of Physics and Astronomy, Louisiana State University, Baton Rouge, LA 70803, USA}
\author{Will M. Farr}
\affiliation{Center for Computational Astrophysics, Flatiron Institute,
162 5th Ave, New York, NY 10010,USA}
\affiliation{Department of Physics and Astronomy, Stony Brook University, Stony Brook NY 11974, USA}
\author{Scott Perkins}
\affiliation{Lawrence Livermore National Laboratory, Livermore, California 94609 USA}
\author{Kaze W. K. Wong}
\affiliation{Center for Computational Astrophysics, Flatiron Institute,
162 5th Ave, New York, NY 10010,USA}
\author{Nicol\'as Yunes}
\affiliation{Department of Physics and Illinois Center for Advanced Studies of the Universe, University of Illinois at Urbana-Champaign, Urbana, IL, 61801, USA}

\begin{abstract}
Extensions to General Relativity (GR) allow the polarization of gravitational waves (GW) from astrophysical sources to suffer from amplitude and velocity birefringence, which respectively induce changes in the ellipticity and orientation of the polarization tensor. We introduce a multi-messenger approach to test this polarization behavior of GWs during their cosmological propagation using binary sources, for which the initial polarization is determined by the inclination and orientation angles of the orbital angular momentum vector with respect to the line of sight. In particular, we use spatially-resolved radio imaging of the jet from a binary neutron star (BNS) merger to constrain the orientation angle and hence the emitted polarization orientation of the GW signal at the site of the merger, and compare to that observed on Earth by GW detectors.  For GW170817, using past measurements of the inclination angle, we constrain the deviation from GR due to amplitude birefringence to $\kappa_A = -0.12^{+0.60}_{-0.61}$, while the velocity birefringence parameter $\kappa_V$ remains unconstrained. 
The inability to constrain $\kappa_V$ is due to the low amplitude of GW170817 in the Virgo detector, and measurements of the polarization orientation require information from a combination of multiple detectors with different alignments. For this reason, we also mock future BNS mergers  with resolved afterglow proper motion and project that $\kappa_V$ could be constrained to a precision of $5\,$rad (corresponding to an angular shift of the GW polarization of $\delta\phi_V\approx 0.2\,$rad for a BNS at $100\,$Mpc) by a future network of third-generation ground-based GW detectors such as Cosmic Explorer and the radio High Sensitivity Array. Crucially, this velocity birefringence effect cannot be constrained with dark binary mergers as it requires polarization information at the emission time, which can be provided only by electromagnetic emission.
\end{abstract}

\date{\today}
\maketitle


\section{Introduction}
In general relativity (GR) the polarization of a gravitational wave (GW) does not vary between emission and detection in a perfectly homogeneous and isotropic expanding universe. However, changes can occur if gravity has non-trivial interactions with an additional cosmological field that breaks chiral symmetry. A common example of a theory that introduces such effects and has been studied in the literature is Chern-Simons (CS) gravity~\cite{Lue:1998mq, Jackiw:2003pm, Alexander:2007kv, Alexander:2009tp}. In theories with such non-trivial and chiral interactions, the cosmological propagation of right- and left-handed polarized GWs can differ since both their relative phases and amplitudes may evolve during propagation, inducing an overall change between the emitted and observed GW polarization. 
These two effects are known as velocity and amplitude birefringence, respectively.

The possibility that GW birefringence could be detected by  space- and ground-based detectors in the context of CS gravity was first analyzed in \cite{Alexander:2007kv, Yunes:2010yf}. These works found that the interferometric response of a GW with amplitude birefringence is partially degenerate with the inclination angle and luminosity distance to the binary system, but that constraints may still be possible. This idea was recently put in practice in~\cite{Okounkova:2021xjv,Wang:2020cub, Zhao:2022pun, Ng:2023jjt, Callister:2023tws}, finding the first constraints on such amplitude birefringence. Reference~\cite{Yunes:2010yf} also argued that coincident GW and $\gamma$-ray burst (GRB) events could be used to break the above degeneracies due to precise localization information obtained from the GRB, leading to an independent constraint. 

In this paper, we propose testing the propagation behavior of GW polarization from merging binary neutron stars (BNSs) with EM counterparts, including amplitude birefringence \cite{Alexander:2007kv, Yunes:2010yf} as well as velocity birefringence. 
In particular, we use radio observations to illustrate the implementation of the idea proposed in \cite{Yunes:2010yf} for amplitude birefringence by using the GW170817 event \cite{TheLIGOScientific:2017qsa} and 
generalize it to velocity birefringence; this extends the scope of past tests of GR performed with this system \cite{LIGOScientific:2018dkp}. In doing so, we also assess the prospects for applying this technique to future BNS events detected by LIGO \cite{LIGO}, Virgo \cite{Virgo}, KAGRA \cite{KAGRA}, and next-generation, ground-based detectors \cite{Evans:2021gyd,LIGOIndia:20211, Saleem:2021iwi}.

The methodology of such a test is as follows. The emitted GW polarization by binary systems in the source frame depends on the direction of emission with respect to the binary's angular momentum vector, and it is hence characterized by two angles: the inclination $\iota$ and orientation (or polarization) $\psi$\footnote{Additional parameters are of course also needed to specify the projection of the emitted GW polarization onto a given detector in the detector frame, such as the location of the source in the sky $(\theta,\phi)$.}
We can infer these two angles by measuring and comparing the amplitude and phase of the signal between different GW detectors.  Separately, multi-band EM detections of the collimated jet released by the merger can be used to constrain the inclination of the binary, while long-term follow-up observations of the jet's afterglow emission can constrain the orientation of the binary's angular momentum by resolving the sky-projected centroid motion of the afterglow. Therefore, these EM constraints provide information on the expected GW polarization at the moment of emission. By comparing the EM information on the source's orientation to the observed GW polarization state, we can therefore constrain the effects of GW birefringence. 

Past constraints on GW birefringence used binary black holes (BBHs) and models in which the polarization change is frequency dependent, thus introducing distortions on the GW chirp predicted by GR \cite{Wang:2020cub, Zhao:2022pun, Ng:2023jjt,Goyal:2023uvm}. 
In this work, we also consider frequency-dependent amplitude birefringence because the frequency running is a generally-expected feature of self-consistent modified gravity theories, such as CS gravity \cite{Lue:1998mq,Jackiw:2003pm,Alexander:2004wk, Alexander:2004us,Alexander:2007kv, Yunes:2010yf}. The constraints we obtain can thus be directly compared to those previously obtained in the literature. In particular, we quantify to what degree the incorporation of additional information from the EM side improve such constraints, versus searches that do not assume an EM counterpart.
Nevertheless, and contrary to previous studies, we do not consider frequency dependence of the velocity birefringence since self-consistent modified gravity theories do not require such a frequency dependence \cite{Jenks:2023pmk}.
This effect does not therefore introduce waveform distortions in our analysis, and instead is completely degenerate with the binary's angular momentum orientation in the GR waveform templates, making it impossible to identify with GW data alone. Hence, only binaries with EM counterparts can constrain such effects, for example, through radio observations that measure the binary orientation and help break the aforementioned degeneracy.  

For the GW170817 event, we obtain a constraint on the amplitude birefringence parameter $\kappa_A=-0.12^{+0.60}_{-0.61}$ at 68\% credibility using all EM information available. This result is weaker than previous BBH constraints \cite{Ng:2023jjt}, mainly due to the fact that BBHs are located much further away than the BNS source of the GW170817 event, and cosmological birefringence grows proportionally with the comoving distance to the source.

For velocity birefringence, GW170817 does not provide meaningful constraints since the GW detectors could not measure the binary's orientation angle. For this reason, we also simulate the measurement of future BNS events, and find that amplitude birefringence can be constrained nearly two orders of magnitude more precisely, with precision $\sigma(\kappa_A)\sim \mathcal{O}(10^{-2})$ at $68\%$ credibility, while the velocity birefringence parameter $\kappa_V$ may be constrained with a precision up to $\sigma(\kappa_V)\sim 5\,$rad with a future network of Cosmic Explorer-like GW detectors, but it might not be possible to obtain meaningful constraints with 2nd generation GW detectors. The main limitation is the fact that velocity birefringence is degenerate with other GW angular parameters, especially for highly face-on/face-off binary systems such as those with observable EM jets. In particular, depending on the binary inclination, we find that the constraints on $\kappa_V$ can be limited by either the GW observations (low inclinations) or by the radio afterglow observations (larger inclinations).

This paper is organized as follows. In Sec.\ \ref{sec:GWbire} we review the polarization of GWs from binary mergers, and introduce the modified cosmological model that incorporates birefringence. Here we discuss how the polarization properties depend on the binary's angular momentum properties and how it can exhibit degeneracies with other angular parameters characterizing the waveform. In Sec.\ \ref{sec:EM} we summarize the EM counterparts of BNS mergers, and how they can be used to measure the expected emitted GW polarization; we also present proof-of-principle results for GW170817 and discuss the prospects for future multi-messenger BNS events. In Sec.\ \ref{sec:results} we explain our methodology for testing birefringence, and show the results for future mock events (like and unlike GW170817); we project observations first assuming a network including LIGO India \cite{LIGOIndia:20211, Saleem:2021iwi} (in addition to the current LIGO, Virgo, and KAGRA), and then Cosmic Explorer \cite{Evans:2021gyd}. In Sec.\ \ref{sec:theory} we translate our birefringence results to constraints on specific fundamental gravity theories that break parity symmetry and that have been previously considered in the literature. Finally, in Sec.\ \ref{sec:conclusions} we summarize our results and discuss future prospects. Throughout this paper we set the speed of light $c=1$, except when reviewing the physics of jet dynamics. 

\section{Birefringent GW Polarization}\label{sec:GWbire}

In this section, we first define the main angles that determine the GW polarization state in compact binary systems. We then introduce amplitude and velocity birefringence, and discuss how they can exhibit degeneracies with GR waveform parameters, which will be confirmed with the numerical results of Sec.\ \ref{sec:results}.

\subsection{Birefringence Review}

GWs in GR and many of its extensions, propagate only two tensor degrees of freedom in metric perturbations---or polarizations---on a cosmological universe\footnote{If the extension to GR incorporates additional degrees of freedom, such as a scalar field, then additional polarizations may be produced at emission, although this is not necessarily the case. In this paper, we assume that those additional polarizations are highly suppressed and hence negligible, as is the case in dynamical CS gravity and in scalar-Gauss-Bonnet gravity~\cite{Wagle:2019mdq}.}. A common basis for these polarizations is one composed of the linear \textit{plus and cross} states, with GW components $h_+$ and $h_\times$, respectively (see, e.g., \cite{Isi:2022mbx} for a review).
The observed detector response $h$ is determined by how the different polarizations interact with the detector and, in the linear basis, can generally be expressed as:
\begin{equation}
    h= F_{+}(\theta,\phi,\psi) h_{+} + F_{\times }(\theta,\phi,\psi)h_{\times},
    \label{strainobs}
\end{equation}
where $F_{+/\times}$ are the antenna pattern functions, which are obtained by taking inner products between tensors representing the detectors and the GW signal.
In detector-centered coordinates, these take the following explicit form for L-shaped GW detectors like LIGO:
\begin{align*}
    F_{+}&= \frac{1}{2}\left[1+\cos^2(\theta)\right]\cos (2\phi) \cos (2\psi) \\
    &-\cos(\theta)\sin (2\phi) \sin(2\psi),\\
    F_{\times} &= \frac{1}{2}\left[1+\cos^2(\theta)\right]\cos (2\phi) \sin (2\psi) \\
    &+\cos(\theta)\sin (2\phi) \cos(2\psi)\, ,
\end{align*}
where $(\theta,\phi)$ are polar and azimuthal angles determining the sky location of the source relative to the detector, and $\psi$ is the orientation angle (also known as polarization angle in the GW literature), orienting the principal axes of the linear polarization frame in the plane of the sky. 

\begin{figure*}[t!]
\centering
\includegraphics[width = 0.9\textwidth]{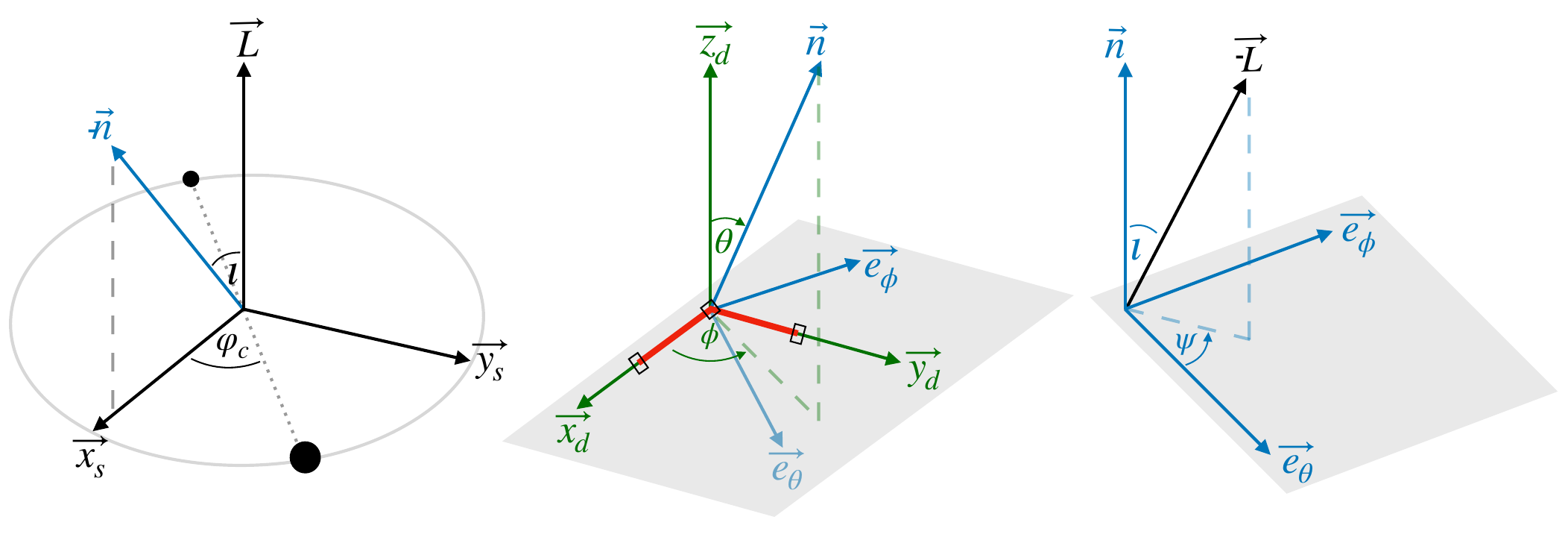}
 \caption{Diagram of source 
(black), detector (green), and sky (blue)  frames
with axes $\{\vec{x}_s,\vec{y}_s, \vec{L}\}$,  $\{\vec{x}_d,\vec{y}_d,\vec{z}_d\}$, and $\{\vec{e}_\theta,\vec{e}_\phi,\vec{n}\}$, respectively; all relevant angles are illustrated.
First, for a non-precessing binary, the source frame is defined with the orbital angular momentum $\vec{L}$ as the $z$ axis and the $x$ axis pointing in some reference direction (typically the ascending node), with respect to which the line from the lightest to the heaviest body defines an angle $\varphi_c$ at a fiducial time (often close to the moment of merger)---this reference phase, $\varphi_c$, is known as the ``coalescence'' phase; additionally, the inclination $\iota$ is the angle between the location of the observer $-\vec{n}$ and $\vec{L}$.
Next, the detector frame is fixed by the L-shaped legs of the detector (shown in red); in this frame, the sky position of the source is prescribed by the angles $\{\theta,\phi\}$, which are the Euler angles that align $\vec{z}_d$ with $\vec{n}$---the line of sight direction from detector to source---and rotate $\{ \vec{x}_d ,\vec{y}_d\}$ to
 $\{ \vec{e}_\theta, \vec{e}_\phi \}$. Finally, in the sky frame,  $\psi$ is a third Euler angle that rotates $\vec{e}_\theta$ to be aligned with the transverse projection of $-\vec{L}$ and describes the orientation of the angular momentum. 
 (For further details on these frames see Ref.~\cite{Isi:2022mbx}.)}
 \label{fig:angles}
\end{figure*}

Figure\ \ref{fig:angles} shows the definitions of the angles characterizing the GW polarization state in the detector frame: $(\theta,\phi)$, $(\iota,\psi)$ and, additionally, a fiducial phase angle $\varphi_c$. This is one of three relevant coordinate frames: source, detector and sky (or wave) frames, which are related to each other by Euler rotations.
Additionally, one often defines the so-called Earth frame to specify the polarization state in a detector-independent way when we have a network of GW detectors. This frame has the origin at the Earth's centroid, its $\hat{z}$ axis pointing towards the north pole and $\hat{x}$ axis towards $(\mathrm{RA},\, \mathrm{Dec})=(0,\,0)$ (and $\hat{y}$ chosen to complete a right-handed coordinate triad). The angles in the Earth frame are defined as in Fig.\ \ref{fig:angles} when the detector frame is replaced by the Earth frame. Then, the response of each detector could be obtained by making appropriate transformations from the Earth frame to the detector frame (see also \cite{Isi:2022mbx} for more details). From now on, we will quote sky location and orientation angles in the Earth frame.

In GR, the two GW polarizations propagate in the same way over cosmological distances, which means that the emitted polarization state will match the detected one, and will be characterized by the angles $(\iota,\psi)$ specifying the location of the observer relative to the source.
In extensions to GR that violate parity gravitationally, the GW polarization can change between emission and detection because the two GW polarizations do not propagate in the same way. 
In such cases, it is convenient to use the alternative basis of \textit{left ($L$) and right ($R$)} circular polarizations, with GW components $h_L$ and $h_R$, which are related to the linear polarization by
\begin{equation}
    h_+=\frac{h_R+h_L}{\sqrt{2}}, \quad  h_\times=i\frac{h_R-h_L}{\sqrt{2}},
\end{equation}
where $h_{L,R}$ are complex quantities, and thus $h_{+,\times}$ are now expressed in complex form.

In parity-violating theories the two circular polarizations are modified with respect to GR.
In the frequency-domain, the observer receives polarizations as \cite{Jenks:2023pmk}
\begin{subequations} \label{eq:birefringence}
\begin{align}
    h_R(f) &= h_R^{\rm GR}(f) \; e^{-\delta\phi_A(f) -i\delta\phi_V(f)},
    \\
    h_L(f) &= h_L^{\rm GR}(f) \; e^{+\delta\phi_A(f) + i\delta\phi_V(f)}
\end{align}
\end{subequations}
where $h_{L,R}^{\rm GR}(f)$ are the frequency-domain, circular polarizations emitted by the source, which we take to be the same as in GR assuming source effects induced by the modified theory are small. Here, the term $\delta \phi_{V}(f)$ (assumed to be a real function) induces a relative phase shift between $L$ and $R$ polarizations, which in turn will produce a rotation of the polarization plane, known as velocity birefringence. In addition, $\delta \phi_A(f)$ (assumed to be a real function) induces a relative amplitude change between $L$ and $R$ polarizations, which will change the polarization ellipticity, known as amplitude birefringence. 

In most cases, BNS systems are well described by a nearly equal-mass binary in a nearly circular orbit. In that case, the $(\ell=2, |m|=2)$ GW angular spherical harmonic dominates the signal, and the polarization predicted by GR during the early inspiral can be approximated in the frequency-domain as \cite{Blanchet:2013haa}:
\begin{align}
    h_{L/R} (f) &= A(f) \left(1 \pm \xi\right)^2 e^{i\left(\tilde{\Psi}(f)+2\varphi_c\right)}, \label{22Pol}
\end{align}
where the $+$ ($-$) sign corresponds to $L$ ($R$), and we have defined $\xi=\cos\iota$.
Additionally, $\tilde{\Psi}(f)$ is the frequency-dependent Fourier GW phase, with $2\varphi_c$ the coalescence phase (see\ Fig.\ \ref{fig:angles}), and $A(f)$ is the Fourier amplitude given by
\begin{equation}
    A(f)\propto \frac{1}{d_L}(G\mathcal{M}_z)^{5/6}(\pi f)^{-7/6},
\end{equation}
where $d_L(z)$ is the luminosity distance to the source, and $\mathcal{M}_z=(1+z)\mathcal{M}^{\rm src}_c$ is the redshifted chirp mass, with $\mathcal{M}^{\rm src}_c=(m_1 m_2)^{3/5}/(m_1+m_2)^{1/5}$ for a binary with individual component masses $m_1$ and $m_2$.

For a given GW detector, the response given by Eq.\ (\ref{strainobs}) in the presence of birefringence is then \cite{Jenks:2023pmk}
\begin{align}
    h &= h^{\rm GR}\left[1+f(F_{+,\times}, \xi)\,\delta\phi_A(f) - g(F_{+,\times}, \xi)\,\delta\phi_V(f)\right] \nonumber\\
    &\times \exp \{ i\left[g(F_{+,\times}, \xi)\,\delta\phi_A(f) + f(F_{+,\times}, \xi)\,\delta\phi_V(f) \right] \}\,, 
    \label{Bire_h}
\end{align}
where we have introduced the auxiliary functions 
\begin{align}
   f(F_{+,\times}, \xi) &= \frac{ 2(F_+^2 + F_\times^2)\xi(1 + \xi^2)}{4F_\times^2\xi^2 + F_+^2(1 + \xi^2)^2} \label{f}, \\
   g(F_{+,\times}, \xi) &= \frac{F_+F_\times(-1 + \xi^2)^2}{4F_\times^2\xi^2 \label{g}
+ F_+^2(1 + \xi^2)^2},
\end{align}
which determine how $\delta\phi_{A,V}(f)$ alters the amplitude and phase of the detector response. This shows that, depending on the sky location, polarization and inclination angles, $\delta\phi_{A,V}(f)$ can affect both the amplitude {\it and} phase of the observed signal, even though $\delta\phi_A(f)$ was defined as an amplitude-only polarization modification and $\delta\phi_V(f)$ as a phase-only polarization modification.

For binaries that are exactly face on/off, we have that $\xi=\pm 1$ and hence $g(F_{+,\times}, \xi)$ vanishes, meaning that $\delta\phi_A(f)$ affects only the amplitude of $h$ and $\delta\phi_V(f)$ affects only the phase of $h$. However, this behavior is not generic since GW events may be arbitrarily oriented.  Nonetheless,  the binary inclination of observed events will be biased towards highly face on/off events due to selection effects \cite{Schutz:2011tw}, and hence, the effects of $g$ will be suppressed for most sources, although potentially detectable with sufficiently high signal-to-noise ratio (SNR).

While the frequency (or time) evolution of the modified effects $\delta \phi_{A,V}(f)$ can exhibit a wide range of possible behavior depending on the gravity theory under consideration \cite{Jenks:2023pmk}, here we focus on the following agnostic yet well-motivated parametrization:
\begin{align}
    \label{eq:par-phiA}
    \delta\phi_A(f)&=\kappa_A \left(\frac{d_c(z)}{1\textrm{Gpc}}\right) \left(\frac{f}{100\textrm{Hz}}\right), 
    \\
    \delta\phi_V(f)& = \delta\phi_V=2\kappa_V \ln(1+z),\label{Bire_def}
\end{align}
where $d_c$ is the comoving distance to the source at redshift $z$ when assuming a cosmology with the best-fit $\Lambda$CDM parameters from Planck 2018 \cite{Aghanim:2018eyx}, $f$ is the detected frequency of the GW, and $\kappa_{A,V}$ are arbitrary constant parameters characterizing the level of parity breaking. While $\kappa_A$ is dimensionless, $\kappa_V$ has units of angle.
By this parametrization, the velocity birefringence introduced by $\delta\phi_V$ is independent of frequency ($\delta\phi_V(f)\equiv \delta\phi_V$).

\begin{figure*}
\centering
\includegraphics[width = 0.95\textwidth]{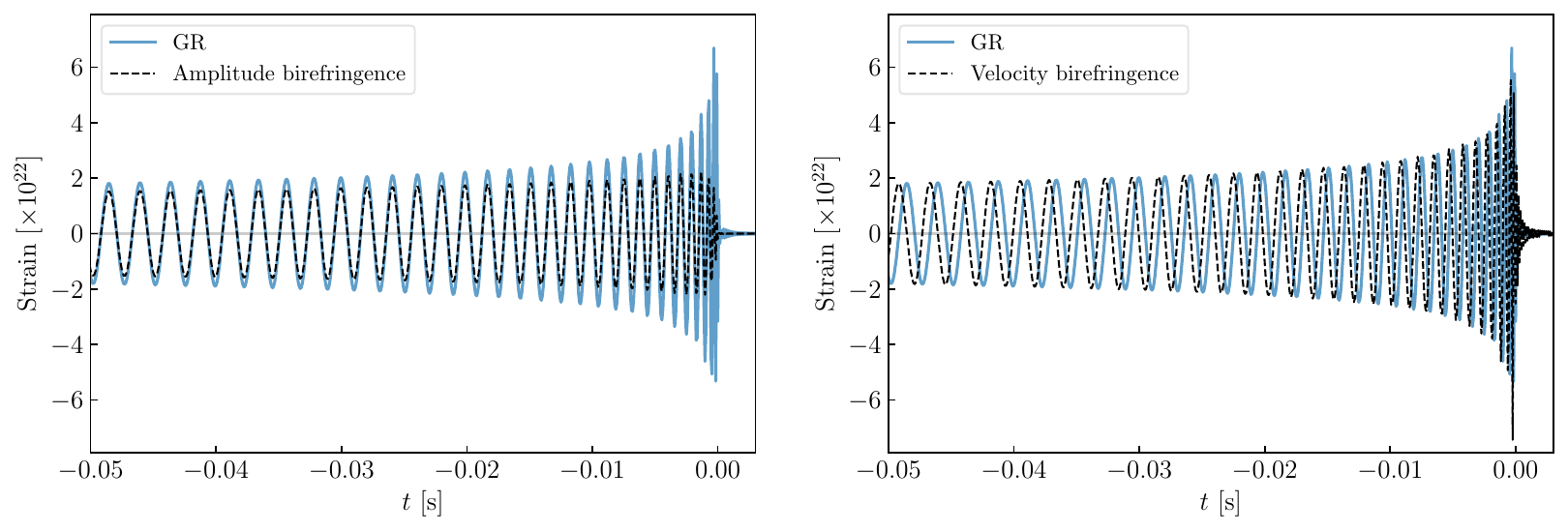}
 \caption{Illustration of the end of a BNS waveform signal, similar to GW170817, with equal masses $m=1.4M_\odot$, redshift $z=0.01$, and inclination $\iota=2.85$rad. Left: we compare the signal expected in GR to that with amplitude birefringence when $\kappa_A=1$. Since this waveform is almost entirely right-handed, we mostly see a suppression in the amplitude that gets enhanced near the merger due to the increase in frequency. Right: we compare the signal expected in GR to that with velocity birefringence when $\kappa_V=-80$rad. In this case, we have  $\delta\phi_V\approx -\pi/2$ and we see a constant shift in the phase during the inspiral. Due to the fact that near the merger there is interference of multiple frequency modes at any given time, the waveform suffers some morphology changes when each frequency mode is shifted by $\delta\phi_V$. }
 \label{fig:BNS_birefringence}
\end{figure*}

The distance and frequency normalization for $\delta\phi_A$ in Eq.\ (\ref{Bire_def}) are chosen conveniently so that, for current GW observations by ground-based detectors, a  shift 
$[\delta\phi_A(f_{\rm high})-\delta\phi_A(f_{\rm low})]$ (where $f_{\rm low,high}$ are the low and high frequency ends of the sensitivity band) of order unity is induced for a $\kappa_A$ of order unity. However, other choices of normalization have also been made in the literature (see e.g.~\cite{Jenks:2023pmk}). Fig.\ \ref{fig:BNS_birefringence} illustrates the phenomenon of amplitude and velocity birefringence in the last few cycles of a BNS waveform, and its comparison to GR.

Since these parity-violating effects arise from propagation over cosmological distances, they are such that $\delta\phi_{A,V}=0$ for sources at $z=0$ (i.e., no deviation from GR), and they grow and accumulate during propagation from higher redshifts. In addition, in self-consistent modified gravity models, $\delta\phi_A$ always depends on odd powers of the GW frequency, while $\delta\phi_V$ depends on even powers \cite{Jenks:2023pmk}. In this paper, we consider the lowest-order terms in frequency, as those provide the leading order effect, which is why $\delta \phi_V$ is a constant and $\delta \phi_A$ is linear in $f$. Formally, the lowest order corrections to $\delta\phi_{A,V}$ applied by \textit{each} both have two terms \cite{Jenks:2023pmk}: one that scales with redshift and one that scales with distance. In this work, however, because we consider only BNS events at very low redshift, these two terms are equivalent, which enables us to use just a single scaling with distance or redshift for each, and which is the same parametrization for $\delta\phi_A$ that was used in \cite{Ng:2023jjt}.  
However, for GW events at larger distances (i.e., when $z \ll 1$ is not valid) the parametrization of Eqs.\ \eqref{eq:par-phiA} and \eqref{Bire_def} is not unique from physical theories, and instead, one should use the more general parametrization introduced in \cite{Jenks:2023pmk}.  

Reference \cite{Jenks:2023pmk} characterized the polarization propagation of GWs under generic conditions, and it is useful to map our parametrization in Eq.\ (\ref{Bire_def}) to theirs.  Translating the parameters, we find that $\kappa_A=\pi 10^{19} \beta_{1_0}$ and 
$\kappa_V=\gamma_{0_0}/4$.\footnote{The difference of many orders of magnitude in $\kappa_A$ occurs because of a different normalization is used. In \cite{Jenks:2023pmk} $c$ is used to normalize $d_c f$, while here we use $1\,$Gpc and $100\,$Hz, which is more convenient for analyzing current detections.}
Furthermore, given the limited detection volume of BNS mergers by current GW detectors, we can use the small-redshift approximation, where $d_c\approx cz/H_0$ and $\ln (1+z)\approx z$. In such cases, the $\kappa_{A,V}$ constraints we obtain can be translated to the more general parameter combination $\kappa_A\approx \pi 10^{19}\left(\beta_{1_0}+\alpha_{1_0}H_0/(\Lambda_\textrm{PV} c)\right)$ and $\kappa_V\approx 1/4(\gamma_{0_0}+\delta_{0_0}c\Lambda_\textrm{PV}/H_0)$. This ``dictionary" allows us to straightforwardly connect $\kappa_{A,V}$ with specific parity-breaking gravity theories studied in the literature, as compiled and derived in \cite{Jenks:2023pmk}. 
As an example, a linear frequency dependence for  $\delta\phi_A$ appears in well-known theories such as dynamical CS gravity \cite{Lue:1998mq, Jackiw:2003pm, Alexander:2007kv, Yunes:2010yf, Alexander:2017jmt} and its variations \cite{Nojiri:2019nar, Sulantay:2022sag}, which predict 
$\alpha_{1_0}\not=0$, while a frequency-independent  $\delta\phi_V$ appears in theories such as Symmetric Teleparallel theories \cite{Conroy:2019ibo}, which predict $\gamma_{0_0}\not=0$. 

While in general the parameters $\kappa_{A,V}$ can take on arbitrary values if we are agnostic about the underlying gravitational interactions, the birefringence gravity models leading to Eq.\ (\ref{Bire_def}) assume that deviations from GR are small, such that $\delta\phi_{A,V}$ depend linearly on $\kappa_{A,V}$.  For this reason, we impose a consistency bound on $\kappa_A$ such that 
\begin{equation}
    |\delta\phi_A|< 1, \label{kA_bound}
\end{equation}
for any frequency within the relevant observable range. In addition, since $\delta\phi_V$ is a periodic phase, we consider an effective range on $\kappa_V$ such that 
\begin{equation}
    \delta\phi_V \in [-\pi,\pi].\label{kV_bound}
\end{equation}
Nevertheless, this phase shift need not be small compared to the GR phase at the observation time. This is because $\delta\phi_V$ is periodic and it is a total phase integrated over the cosmological propagation time, and instead it is enough to assume small \textit{local} deviations from GR in order to match this parametrization to modified gravity theories. 
The numerical analyses in Sec.\ \ref{sec:results} will use these conditions to set appropriate priors on $\kappa_{A,V}$.

\subsection{Degeneracies}

\subsubsection{Velocity Birefringence}\label{sec:velbire}

For the specific parametrization considered in this paper, $\delta \phi_V$ is a constant, so it consists of a simple phase shift between $L$ and $R$ polarizations. In this case, Refs.~\cite{Ezquiaga:2021ler, Isi:2022mbx} show that this effect is exactly degenerate with a shift in the orientation angle $\psi$ for one or multiple detectors. Indeed, by making the redefinition 
\begin{equation}
\psi\rightarrow \psi + \delta\phi_V/2\, , \label{Psi_degener}
\end{equation}
the effect of velocity birefringence is reabsorbed into the orientation angle in $F_+$ and $F_\times$ (cf.~Eq.~\eqref{eq:birefringence} here to Eq.~(38) in \cite{Isi:2022mbx}).
This means that we can define two separate values of $\psi$, at the moment of emission and detection, related by:
\begin{equation}
    \psi^{\rm det}= \psi^{\rm em} + \delta\phi_V/2\, . \label{Psi_degener2}
\end{equation}
Therefore, the only way to constrain $\kappa_V$ is by having two independent observations providing $\psi^{\rm em}$ and $\psi^{\rm det}$. For this reason, frequency-independent velocity birefringence can only be probed with multi-messenger events in which the EM observations allow us to reconstruct $\psi^{\rm em}$, while the GW observations measure the final orientation of the received waves, $\psi^{\rm det}$. Indeed, as we will discuss later, for BNS mergers with EM counterparts, radio observations of the afterglow jet can constrain the orientation of the vector $\vec{J}$ on the sky, and hence they can constrain $\psi$ at emission (cf.\ Fig.\ \ref{fig:angles}). 
We also note that this type of modified effect cannot be probed with a population of BBHs either, because their orientation distribution is expected to be isotropic with or without velocity birefringence. 

We emphasize that the exact degeneracy in Eq.\ (\ref{Psi_degener}) only occurs because we have assumed $\delta\phi_V$ to be frequency independent. This comes from modified propagation equations for $L$ and $R$ handed polarizations, with a dispersion relation of the form:
\begin{equation}
    \omega^2_{L,R}= k^2 + \epsilon \lambda_{L,R} k, \label{eq:DR}
\end{equation}
where $\lambda_{L}=+1$ and $\lambda_{R}=-1$, and $\epsilon=0$ recovers the usual dispersion relation in GR. Since we are assuming that the deviations from GR are small, then Eq.\ (\ref{eq:DR}) does not lead to any change in the group velocity of GWs (linearly in $\epsilon$), i.e., $v_g \sim 1 + [\lambda_{L,R} \epsilon/(8 \kappa)]^2$, and hence no physical time delay and waveform distortion will be present. Instead, $\phi_V$ only leads to a frequency-independent phase shift, as assumed in Eq.\ (\ref{Bire_def}).\footnote{This is contrary to the previous studies in \cite{Qiao:2019wsh, Wu:2021ndf}, which consider Eq.\ (\ref{eq:DR}) and calculate a frequency-dependent time delay using the phase velocity of the wave, which leads to a distortion of the signal that can be constrained without EM counterparts. This use of phase velocity is not appropriate, since the phase velocity does not determine the physical speed of GW propagation, as discussed in \cite{Ezquiaga:2022nak}.}

An additional frequency dependence in $\delta\phi_V$ would break the aforementioned degeneracy and induce observable phase distortions in the waveform. This is the case for velocity birefringence previously studied in \cite{Wang:2020cub, Gong:2021jgg, Zhao:2022pun}, for which phase distortions can be probed using any individual binary event, even those without EM counterparts. Nonetheless, from an effective field theory point of view, higher-order frequency terms are further suppressed by the theory cut-off scale and thus are naturally expected to be small \cite{Jenks:2023pmk} in self-consistent gravity theories. Furthermore, from a practical point of view, these frequency-dependent corrections would appear at least at $f^2$ order in $\delta\phi_V$, which means that  they are sub-dominant during the early inspiral of binaries, and only near-merger data could provide meaningful constraints on such corrections, which could additionally be contaminated by waveform mismodeling systematics around the merger.

Other angular parameters in GR, such as the coalescence phase $\varphi_c$, also introduce frequency-independent phase shifts to the waveform and could potentially induce degeneracies with $\delta\phi_V$. As shown by Eq.\ (\ref{22Pol}), $\varphi_c$ enters both $L$ and $R$ polarizations in the same way and hence does not introduce a relative phase shift, contrary to $\delta \phi_V$.  Nevertheless, when a binary is viewed precisely face-on or face-off (i.e., $\cos\iota=\xi=\pm 1$), it will emit a purely circular GW polarization, and $\varphi_c$ will be exactly degenerate with $\delta\phi_V$, for any number of GW detectors.
Instead, for binaries viewed from larger inclination angles, this degeneracy breaks, such that in the extreme case of perfectly edge-on system (i.e., $\cos\iota=0$), even a single detector can distinguish the two angles.  

For unequal mass binaries in a nearly circular orbit, the GW signal can receive significant contributions from additional angular spherical multipoles $(\ell,m)$ beyond just the $(2,2)$ mode.  These can in principle help break degeneracies between $\varphi_c$ and $\delta\phi_V$ because $\varphi_c$ generally enters the phase of each mode as $m\varphi_c$ in both $L$ and $R$ polarizations \cite{RevModPhys.52.299, Kidder:2007rt}.  However, for binaries viewed exactly face on/off, the GW signal again contains only $(\ell, |m|=2)$ harmonics, regardless of mass ratio, and the degeneracies previously discussed will still hold.

The fact that face-on/off binaries will be subject to degeneracies between $\varphi_c$ and $\delta\phi_V$ for circular binaries regardless of the binary properties may present a problem for using BNS with EM counterparts. This is because the binary orbital angular momentum needs to be highly aligned with the line of sight to observe a bright $\gamma$-ray burst and associated afterglows. Nevertheless, since most binaries will still possess significant inclination angles, high sensitivity future GW detectors, such as 3G detectors, will be able to break this degeneracy.

Our discussion thus far has focused on how $\delta\phi_V$ can exhibit degeneracies with the angular parameters of GR waveforms. However, from Eq.\ (\ref{Bire_h}) we see that, if $g\not=0$ (i.e., $\iota \neq 0$ or $2\pi$), then $\delta\phi_V$ also affects the observed amplitude. For instance, in the case of GW170817, the EM observations discussed in Sec.\ \ref{sec:EM} provide sky location, polarization and inclination constraints, such that $f(F_{+,\times},\xi)\sim \mathcal{O}(1)$ and $g(F_{+,\times},\xi)\sim \mathcal{O}(10^{-4})$ for the LIGO detectors. Therefore, as we will see in Sec.\ \ref{sec:results}, $g\not=0$ will result in $\kappa_V$ exhibiting mild degeneracies with amplitude parameters, such as $d_L$ for events with sufficiently high SNR.

\subsubsection{Amplitude Birefringence}

The effect of amplitude birefringence can also exhibit degeneracies with GR waveform parameters. If $\delta\phi_A$ was frequency independent, then it would be mostly degenerate with the inclination angle $\iota$ \cite{Alexander:2007kv} for simple waveforms dominated by a quadrupole angular harmonic $\ell=|m|=2$, as in that case (cf.\ Eq.~\ref{22Pol}):
\begin{equation}
    \frac{|h_L^{GR}|}{|h_R^{GR}|}\approx \left(\frac{1+\xi}{1-\xi}\right)^2.
\end{equation}
Since $\delta\phi_A$ changes the overall amplitude of the polarizations, then a corresponding change in the observed luminosity distance of the GWs will be additionally introduced. 
Such a frequency-independent amplitude birefringence can be identified using a catalog of GW sources through departures from isotropy in the inferred orientations of sources \cite{Okounkova:2019dfo,Vitale:2022pmu,Isi:2023dlk}.
Additionally, as we will discuss later, EM counterparts can constrain $\iota$ and $z$, breaking these potential degeneracies, as first suggested in~\cite{Yunes:2010yf}. 

When taking into account frequency dependence (as expected from self-consistent parity-breaking theories, and as assumed in this paper), there will be no exact degeneracy between $\delta\phi_A$ and any GR parameter. In fact, the frequency dependence will induce amplitude distortions to the GR waveform that can be observable and thus constrained by individual BBH events, as done in \cite{Ng:2023jjt}. In some cases, especially for short signals, frequency-dependent amplitude birefringence may nonetheless be (partially) degenerate with effects such as precession, as shown in \cite{Ng:2023jjt}.  In this paper, we will search for these amplitude distortions in BNS systems.

Finally, as also discussed in Ref.\ \cite{Ng:2023jjt}, we emphasize that $\delta\phi_A$ can also affect the phase of the observed detector response, according to Eq.\ (\ref{Bire_h}), which could introduce approximate degeneracies with other GW parameters, as with spins. In fact, since we assume $\delta\phi_A$ scales linearly with frequency, in Sec. \ref{sec:results} we show that it can exhibit degeneracies with the so-called coalescence time $t_c$. This parameter controls the time of arrival (i.e., the placement of the signal in a segment of data), and it enters the frequency-domain phase of the GW signal linearly with frequency: $\tilde{\Psi}(f) \supset f t_c$ in Eq.\ (\ref{22Pol}). Nevertheless, this degeneracy will be broken with a network of GW detectors. 

\section{EM constraints on the orientation angle at emission}\label{sec:EM}

\subsection{Methodology and GW170817}

Binary neutron star mergers can produce EM emissions spanning nearly the entire wavelength spectrum, as was the case for LIGO-Virgo's first BNS merger, the GW170817 event \cite{LIGOScientific:2017zic, LIGOScientific:2017ync}. These observations can be used to localize the source three-dimensionally, placing it in the sky through $(\mathrm{RA}, \mathrm{Dec})$ and providing a measurement of its distance $d_L$ or redshift $z$ (provided the host galaxy is identified), as well as to characterize the orientation of the system through the angles $(\iota,\psi)$.  

\begin{figure*}[t!]
\centering
\includegraphics[width = 0.80\textwidth]{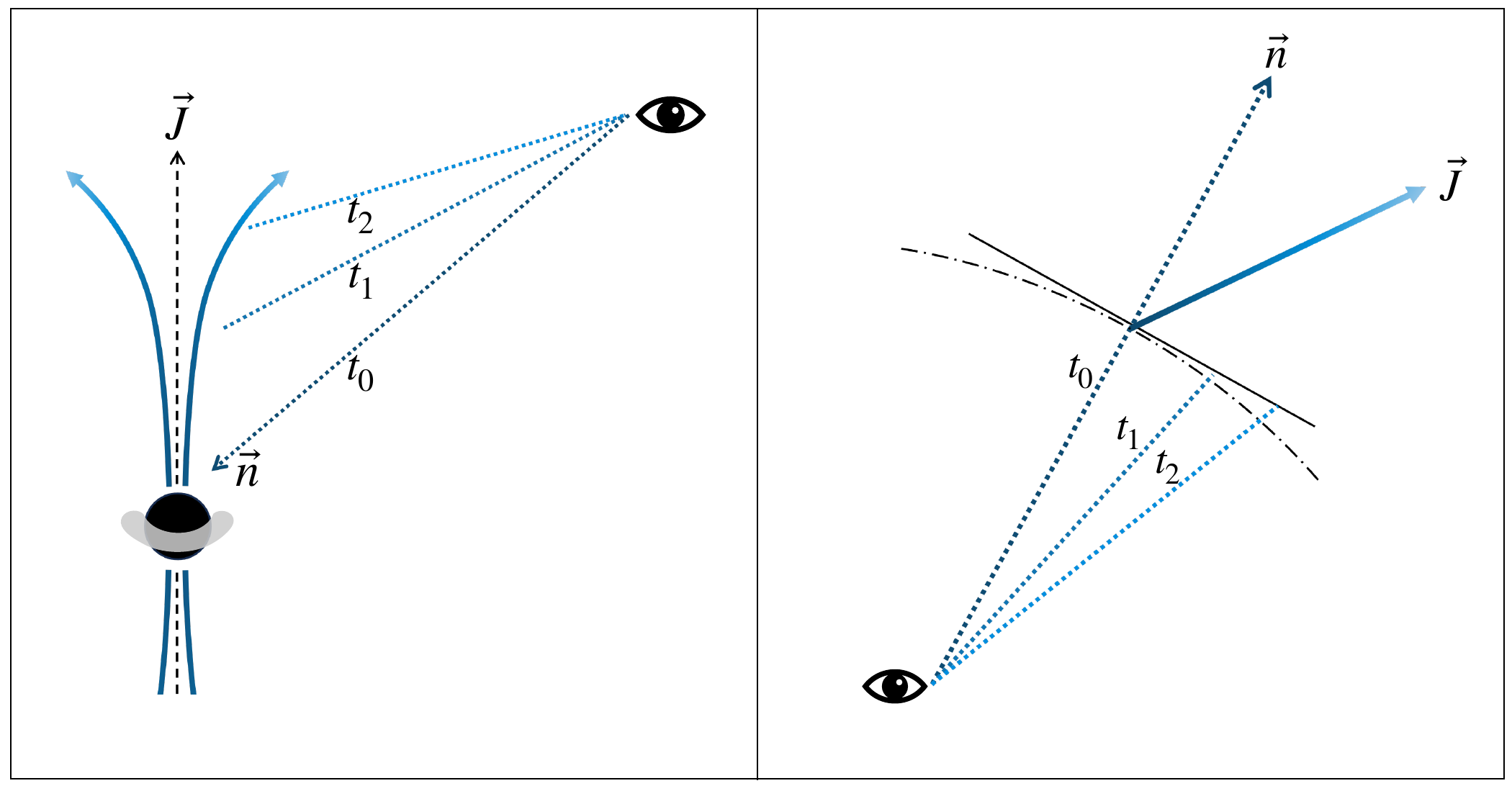}
 \caption{Toy illustrations of the BNS collimated jet and the observed afterglow displacement. Left: the remnant forms an accretion disk that powers a collimated jet (shown in blue shades) along the rotation axis given by the angular momentum vector $\vec{J}$. As time goes on, the jet propagates away from the remnant and an off-axis observer sees the afterglow source getting displaced between different times $t_0<t_1<t_2$. When $t_0$ corresponds to the time of merger, the observation direction is given by the vector $\vec{n}$, which determines the line of sight for GW observations (cf.\ Fig.\ \ref{fig:angles}).
 Right: Point of view of the observer. At some location in the sky, at time $t_0$ a jet along $\vec{J}$ is emitted, and as the jet propagates in time, the observer sees a displacement of the EM source. The sky coordinates of this displacement indicate the direction of the afterglow motion, which is determined by the projection of $\vec{J}$ onto the sky plane perpendicular to $\vec{n}$. This afterglow direction allows to reconstruct the orientation angle $\psi$ (cf.\ Fig.\ \ref{fig:angles}).}
 \label{fig:BNSJet}
\end{figure*}

In particular, BNS mergers produce narrowly collimated  relativistic jets, which 
are naturally expected to propagate along the binary's total
angular momentum $\vec{J}$ (which we assume to be dominated by the binary's orbital angular momentum $\vec{L}$, since BNS spins are expected to be small). 
As a result of the interaction of the front edge of a jet with the interstellar medium (ISM), the jets produce long-lasting synchrotron emission across 
multiple wavelengths, which is referred to as the ``afterglow''. As the jet's front continues to propagate, the afterglow-emitting region moves.
When the viewing angle $\theta_v$ (equivalent to $\iota$ or $\pi-\iota$, depending on whether the binary system is face-on or face-off) is larger than the jet's half opening angle $\theta_j$, high-angular resolution measurements, such as Very Large Baseline Interferometer (VLBI) observations, can resolve the jet's proper motion and hence trace its trajectory. This allows one to infer the orientation angle $\psi$ of the binary's angular momentum (cf.\ Fig.\ \ref{fig:angles}) without detailed afterglow modeling. A toy illustration of this situation is shown in Fig.\ \ref{fig:BNSJet}.

For instance, Ref.~\cite{Mooley:2018qfh} used high-angular resolution VLBI measurements to follow the centroid motion of the radio signal emitted from GW170817, from $75$ to $230$ days after merger. During this time, they observed a significant displacement in RA of 
$2.67\pm0.19\pm0.21$ milliarcsec (mas), but no displacement (within observational uncertainties) in Dec $0.2\pm 0.6\pm 0.7$ mas, where the first and second $1\sigma$ uncertainties correspond to statistical and systematic errors, respectively. From these observations, we will constrain $\psi = 3.14\pm 0.09\,$rad at $68\%$ credibility.

\begin{figure}[h!]
\centering
\includegraphics[width = 0.5\textwidth]{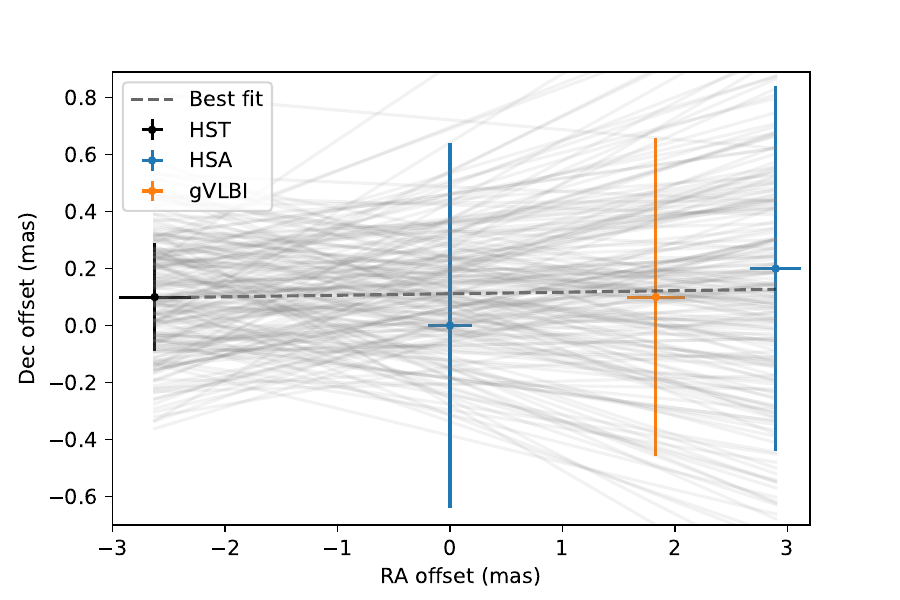}
 \caption{Linear fit to sky offset  locations of GW170817 EM counterparts, when considering 3 radio data points from the afterglow \cite{Mooley:2018qfh, Ghirlanda:2018uyx} and 1 optical point from the kilonova \cite{Mooley:2022uqa}, assumed to represent the true merger location.  Multiple grey solid lines show draws from the posterior of possible linear fits, while the grey dashed line shows the median fit. From this, we constrain the binary's angular momentum orientation angle to be $\psi= 3.14\pm 0.09$ rad at 68\% CL. This is calculated as $\pi$ minus the angle between the horizontal and the fitted line, since $\psi$ is measured from the negative-RA direction in the LIGO convention \cite{Isi:2022mbx}.}
 \label{fig:EMPsi2}
\end{figure}

A more complete set of data points is shown in Fig.\ \ref{fig:EMPsi2}, where we combine optical \cite{Mooley:2022uqa} and radio \cite{Mooley:2018qfh, Ghirlanda:2018uyx} data, indicating the location of the centroid EM emission at different epochs in the Earth frame.\footnote{We note that the radio data in \cite{Mooley:2018qfh, Ghirlanda:2018uyx} was reanalyzed in \cite{Mooley:2022uqa} so for Fig.\ \ref{fig:EMPsi2} we use the latest results summarized in Table 4 in \cite{Mooley:2022uqa}, including statistical and systematic errors.} Following conventions in the literature, the origin is set to be the central value of the first epoch of radio observation: (RA, Dec) = (13:09:48.068638, $-$23:22:53.3909) \cite{Mooley:2018qfh}, taken 75 days after the merger.  However, the true binary location is closer to the position of the optical emission, which is dominated by the kilonova emission from more slowly expanding merger ejecta (black point in Fig.\ \ref{fig:EMPsi2}). 
Figure\ \ref{fig:EMPsi2} shows a linear fit to the jet's trajectory in the sky, from which we infer the binary's orientation angle $\psi$, the median of which is shown as a dashed line.  With respect to the Earth frame, we obtain $\psi=3.14\pm 0.09$ rad. 

In determining the orientation angle this way, the only assumption we have made is that the jet is launched along $\vec{J}\approx \vec{L}$. This differs from most constraints on the inclination angle $\iota$ from EM observations, which require detailed modeling of the jet structure and its associated uncertainties \cite{Lamb:2021use,Ryan:2023pzk}. 
In particular, it is possible to constrain the viewing angle $\theta_v$ (and hence $\iota$) with  afterglow  observations, if information on the time evolution of the flux and centroid position is incorporated.
This is because the displacement and time at which the flux peaks occur contain direct information about the viewing angle $\theta_v$. 
In order to illustrate this, let us summarize some important concepts of the motion of 
relativistic jets (see, e.g., \cite{Eerten2010ApJ,Govreen-Segal2023} for a detailed discussion of the jet dynamics). 

In neutron star mergers, relativistic jets are launched on a short time scale $\lesssim 1$ sec, while the afterglow phase lasts for longer time scales---ranging from days to years. Therefore, 
the jet during the afterglow phase can be considered as an adiabatically-expanding shock wave, with the radius of the shock from the site of the merger given by \citep{Blandford1976PhFl...19.1130B}  
\begin{align}
    R & \approx \left(\frac{17E_{\rm iso}}{16\pi m_p n \Gamma^2 c^2} \right)^{1/3},\\
    & \approx 8\cdot 10^{18}\,{\rm cm}\; E_{\rm iso,52}^{1/3}n_{-3}^{-1/3}(\Gamma/3)^{-2/3},
\end{align}
where $\Gamma \equiv (1-\beta^{2})^{-1/2}$ is the bulk Lorentz factor of the gas shocked by the jet of (normalized) radial velocity $\beta = v/c$, $n_{-3}$ is the density of the interstellar medium in units of $10^{-3}\text{cm}^{-3}$,  $m_p$ is the proton mass, $E_\text{iso}$ is the isotropic-equivalent energy of the shocked gas, and $E_\text{iso,52}$ is the energy $E_\text{iso}$ normalized by $10^{52}\,{\rm erg}$. As this shock propagates outwards, its radius grows and it decelerates with observer time $t$ as $R \propto t^{-3/8}$, until eventually becoming Newtonian, once $\Gamma \sim 1$.

The angular distance to the jet from the merger location for an off-axis observer at a viewing angle $\theta_v$ larger than the jet's half opening angle $\theta_j$, i.e., $\theta_v \gg\theta_{j}$, is given by
\begin{align}
    \delta \theta & \approx \frac{R \sin \theta_v}{d_A},
\end{align}
where $d_A$ is the angular diameter distance to the merger.  Because of relativistic beaming of the emission by the bulk motion of the shocked gas into an opening angle of $1/\Gamma$, the afterglow light curve peaks when $\Gamma$ reaches the value $\Gamma_{\rm peak} \approx (\theta_v-\theta_j)^{-1}$. Therefore, the angular distance 
at the time of the light curve peak for small jet and observing angles such that $\theta_v\gg \theta_j$ is 
\begin{align}\label{peak-offset}
    \delta \theta_{\rm peak} & \approx 2\,{\rm mas\,}E_{\rm iso,52}^{1/3}n_{-3}^{-1/3}\left(\frac{\Gamma_{\rm peak}}{3}\right)^{-5/3}\left(\frac{d_A}{100\,{\rm Mpc}}\right)^{-1},
\end{align}
which depends on $\theta_v$, through $\Gamma_{\rm peak}$. 
In addition, the apparent velocity due to the proper motion of the jet around the peak is roughly $\beta^{\rm app} \approx \Gamma_{\rm peak} \,\beta(\Gamma_{\rm peak})$, so that $\theta_v$ also affects the time at which the flux peak occurs.
As we can see here, other parameters, such as $\theta_j$, also affect the observations at the peak, and this is why the full time evolution of the afterglow has to be carefully incorporated in the analysis in order to measure all of the jet parameters, as done in \cite{Hotokezaka:2018dfi}.
This method was used to improve constraints on the inclination angle of the BNS system GW170817 to $\iota= 2.85 \pm 0.03\, $rad, and thus, on the Hubble constant $H_0$ \cite{Hotokezaka:2018dfi}.

\subsection{Future measurements}

Next, let us estimate how the uncertainties on $\psi$ are expected to scale for future BNS events. The angular resolution of VLBI observations depends on the detector sensitivity, configuration, and source flux (see \cite{Dobie2020MNRAS.494.2449D} for a discussion of the detectability of the jet motion in future BNS mergers, and for the capabilities of current radio facilities).  The astrometric angular accuracy of VLBI observations can be estimated as
\begin{align} \label{eq:theta-res}
    \theta_{\rm res} = \sqrt{\theta_{\rm sys}^2 + \theta_{\rm st}^2}\, ,  
\end{align}
where $\theta_{\rm sys}$ and $\theta_{\rm st}$ are the systematic and statistical uncertainties, respectively, which will determine directly the precision we can achieve on $\psi$.  The statistical component is given by
\begin{align}\label{thetaST}
    \theta_{\rm st} = \frac{\theta_{B}}{\rho_{\rm EM} \sqrt{8\ln 2}}\, ,
\end{align}
where $\theta_B$ is the VLBI beam size and $\rho_{\rm EM}$ is the EM signal-to-noise ratio, which, in turn, is directly proportional to the source flux $F_{\nu,{\rm peak}}$.
In order to estimate $\rho_{\rm EM} \propto F_{\nu,{\rm peak}}$, we can use the fact that the peak radio flux scales according to \cite{Nakar2021}
\begin{align}\label{peakF}
    F_{\nu\,{\rm peak}} \propto E_{\rm iso}n^{\frac{p+1}{4}}\epsilon_e^{p-1} \epsilon_B^{\frac{p+1}{4}}
    \theta_v^{-2p}d_L^{-2},
\end{align}
where $\epsilon_{e}/\epsilon_{B}$ are microphysical parameters that represent the fraction of the post-shock energy placed into relativistic electrons and magnetic fields, respectively; $p$ is the power-law index of the electron energy distribution $dN/dE \propto E^{-p}$; and $\theta_v$ is the viewing angle.  For parameters appropriate for the GW170817 event (i.e., those that match the observed afterglow light curve and the superluminous motion \cite{Hotokezaka:2018dfi}), corresponding to jet properties $E_{\rm iso}=10^{52}\,{\rm erg}$, $\theta_{j}=0.05\,{\rm rad}$ and
$(\epsilon_e,\,\epsilon_B,\,p)=(0.1,\,0.01,\,2.16)$, we find:
\begin{align} \label{eq:flux}
    F_{\nu\,{\rm peak}} \approx 100\,{\rm \mu Jy}\left(\frac{\theta_v}{0.25\,{\rm rad}}\right)^{-4.3}\left(\frac{d_L}{40\,{\rm Mpc}}\right)^{-2}.
\end{align}
Following \cite{Dobie2020MNRAS.494.2449D}, in what follows
we consider the High Sensitivity Array (HSA), consisting of the Green Bank Telescope, the VLA, and the Very Long Baseline Array, an
$S_{\rm noise}=3.2\,{\rm \mu Jy}$ for a 2 hr observation,  $\theta_B=3\,{\rm mas}$ and $\theta_{\rm sys}=0.09\,{\rm mas}$, to study the measurability of the orientation angle $\psi$.

The angular offset of a jet from the merger site can be measured by comparing the jet position obtained from VLBI with the kilonova position. Here we assume that the merger location is determined with the kilonova position from the James Webb Space Telescope (JWST), which can measure the position to a precision of only $\sim 0.05\,{\rm mas}$ for an event at $100\,{\rm Mpc}$ \cite{Mooley:2022uqa}. A discussion on the systematic uncertainties associated with these observations can be found in \cite{Mooley:2022uqa}.

\begin{figure}[h!]
\centering
\includegraphics[width = 0.44\textwidth]
{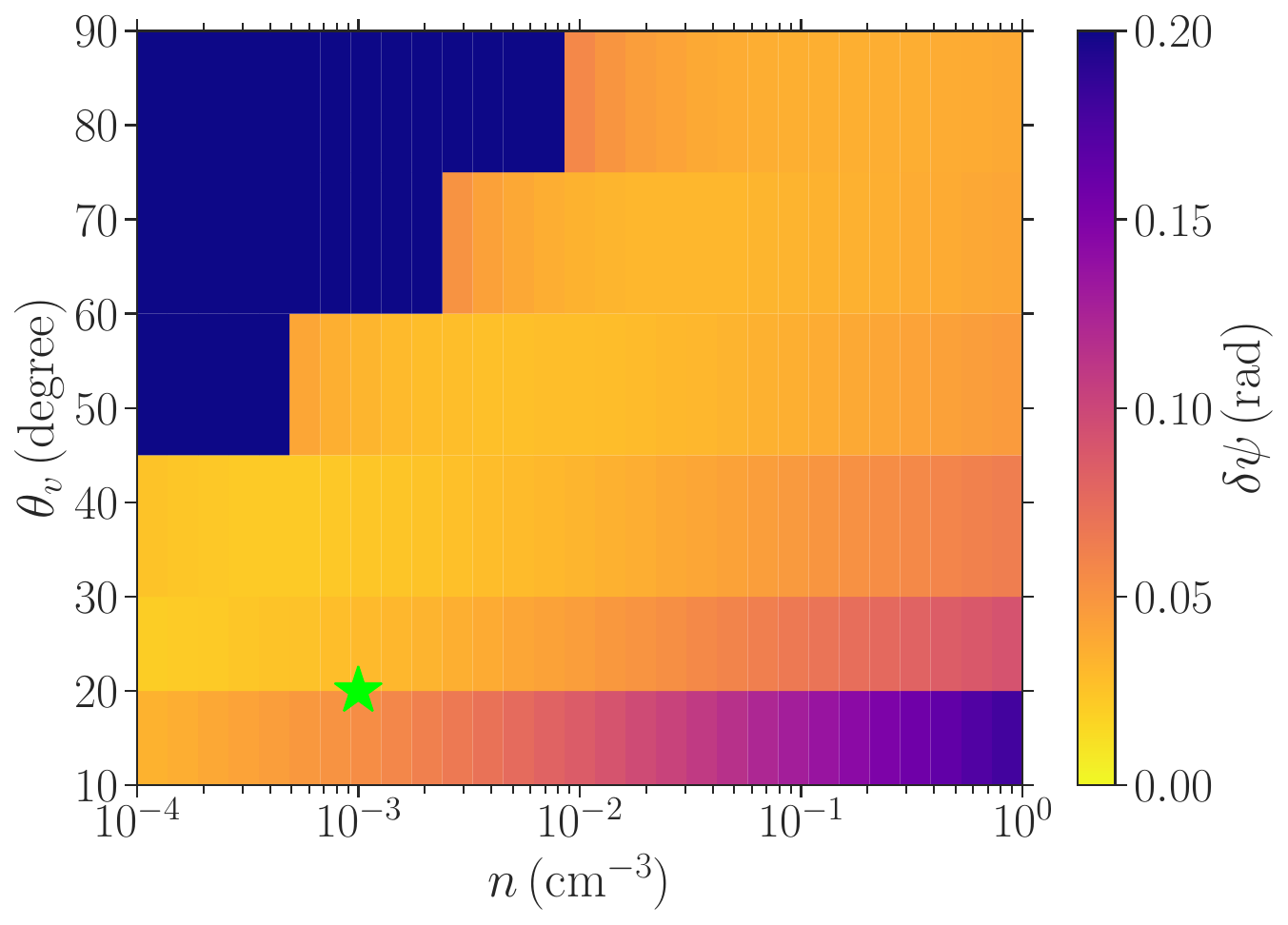}\\
\includegraphics[width = 0.44\textwidth]{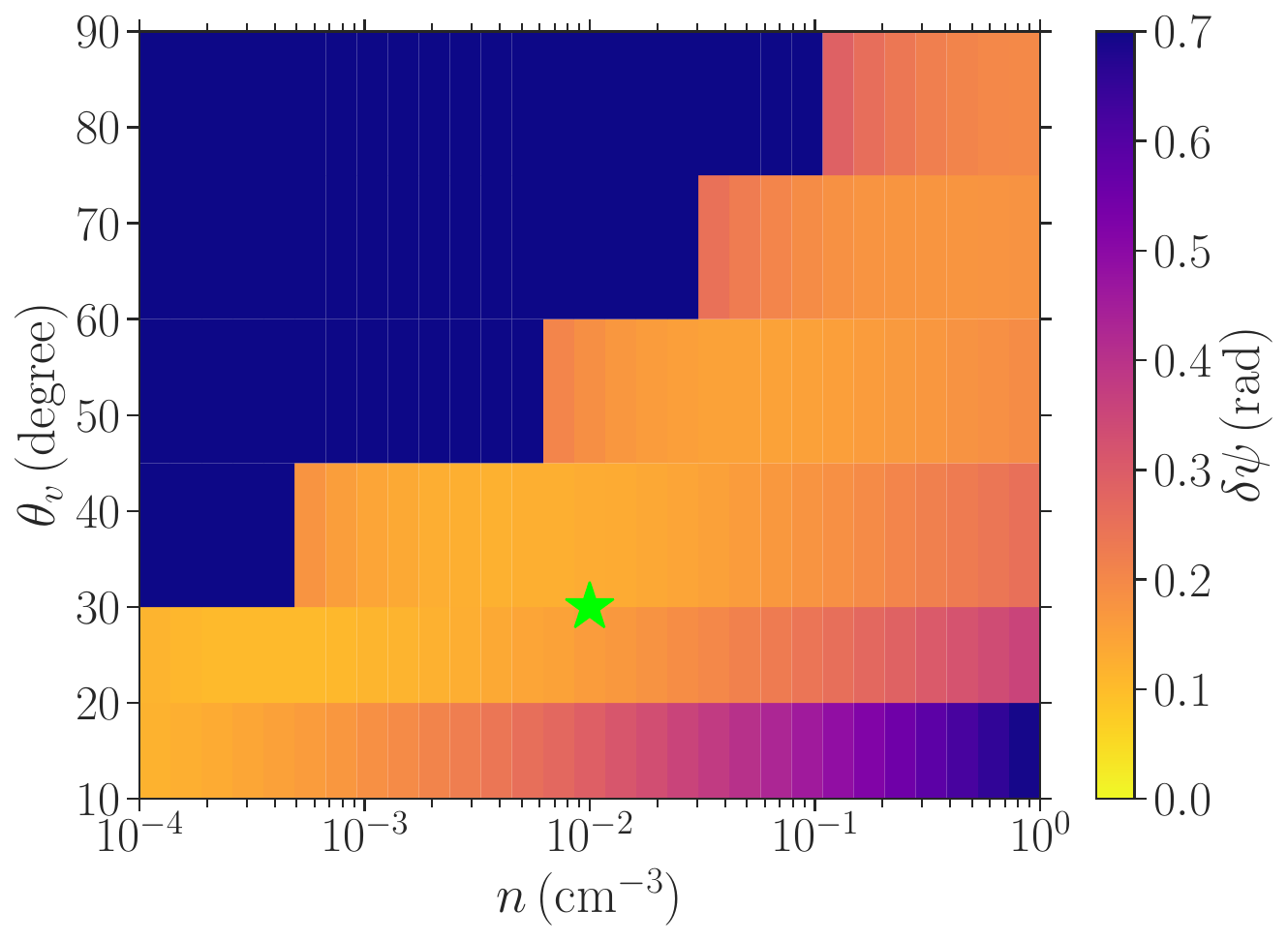}
 \caption{The expected $1\sigma$ uncertainty in the orientation angle measurement of the jet motion with HSA, as a function of the observation angle and interstellar medium density, for a BNS merger at a distance of $40\,{\rm Mpc}$ (upper) and $100\,{\rm Mpc}$ (lower) with an integration time of $2$ hrs. Here we assume that the VLBI observations are capable to measure the source location if the radio flux is detectable with a high significance, $\geq 5\sigma$. We also assume that the merger location is precisely determined by kilonova observations with JWST \cite{Mooley:2022uqa}. Stars show the parameters used for the two mock events analyzed later: a  GW170817-like BNS and a representative BNS. }
 \label{fig:theta_n}
\end{figure}

Figure \ref{fig:theta_n} shows the expected precision $\delta\psi$ of the measurement of the orientation angle from the jet motion for a future merger event at a distance of 40 Mpc (top panel) and 100 Mpc (bottom panel), as a function of the viewing angle $\theta_v$ (interchangeable with binary inclination $\iota$) and the interstellar medium density $n$.  We have here again assumed the GW170817 jet values $E_{\rm iso}$, $\theta_{j}$ and 
$(\epsilon_e,\,\epsilon_B,\,p)$. 
As expected, the VLBI observations are capable of measuring the orientation angle to better than a fraction of a radian for sufficiently small inclination angles and high interstellar medium densities (the blue region in Fig.~\ref{fig:theta_n} corresponds to where the radio flux is too faint to be detected by VLBI observations). The expected $1\sigma$ uncertainty in the orientation angle is typically of the order of 0.1 rad if the motion is detectable. Figure\ \ref{fig:theta_n} highlights with stars the two mock events that will be studied in more detail in Sec.\ \ref{sec:results}, which correspond to a GW170817-like event (upper panel) and a more representative and common BNS event (lower panel).

Currently, the measurement uncertainties on $\psi$ are dominated by the systematic uncertainty $\theta_{\rm sys}=0.09\,{\rm mas}$.  If the latter can be reduced by a factor of a few with improved VLBI observations of a future event, the constraints obtained on the orientation angles can be improved to $\delta \psi \lesssim 0.1\,{\rm rad}$. For example, the next generation Very Large Array (ngVLA) which includes a long baseline array could improve the sensitivity by a factor of $\sim 10$ while also forming a reduced beam size \cite{Selina2018ASPC}. Thus, the ngVLA may provide much greater capabilities to resolve the jet motion in the future.

While birefringence effects accumulate linearly with distance, the precision on $\psi$ allowed from EM observations degrades linearly or cubically with distance, depending on whether the systematic or statistical errors dominate, respectively. This is because the $\psi$ precision depends on how well the afterglow offset $\delta\theta/\theta_{\rm res}$ can be measured, and from Eq.\ (\ref{peak-offset}) we see that $\delta\theta\propto 1/d_L$ while $\theta_{\rm res}$ scales as $d_L^{-2}$ due to statistical effects according to Eq.\ (\ref{thetaST}) when $\theta_{\rm sys}>\theta_{\rm st}$. For this reason, nearby events are expected to be better candidates for increasing the EM measurement precision. 

Finally, we emphasize that in order to observe emission from the collimated jet or its afterglow, the binary's total angular momentum must be relatively tightly aligned with the line of sight.  However, the kilonova emission (powered by radioactive decay of heavy elements expanding at non-relativistic speeds \cite{Metzger+10}) is comparatively isotropic, making it visible regardless of inclination angle and hence likely to be detected for a greater fraction of mergers \cite{Chen:2020zoq}. While kilonova observations are unlikely to tightly constrain the viewing angle of the binary, they can provide accurate source localization measurements. For the GW170817 event, the source was localized at $({\rm RA, Dec})= (13{:}09{:}48.085 \pm 0.018, -23{:}22{:}53.343\pm 0.218)$ \cite{LIGOScientific:2017ync} from the kilonova. Follow-up observations of the galaxy host NGC 4993 provided the distance measurement $d_L=42.9\pm 3.2\,$Mpc \cite{LIGOScientific:2017zic}. As we will see later, having these measurements can help constrain birefringence better as it helps break mild parameter degeneracies.

\section{Birefringence Polarization Test}\label{sec:results}

In this section, we begin in Sec.~\ref{sec:method} by describing the methodology we will use to perform the numerical analysis of cosmological birefringence. In Sec.~\ref{sec:170817} we discuss the results on amplitude and velocity birefringence using the event GW170817. Finally, in Sec.~\ref{sec:mockBNS} we discuss how these constraints will improve with future BNS multi-messenger detections. 

\subsection{Methodology}\label{sec:method}

In order to obtain constraints on $\vec{\kappa}=(\kappa_{A},\kappa_V)$ with real or mocked data, we perform parameter estimation (PE) in the Bayesian inference framework, using either only the GW data, $D_{GW}$, or the GW data in combination with EM data, $D_{EM}$.
Letting $\vec{\xi}$ be the set of waveform parameters other than $\vec{\kappa}$ (e.g., masses, spins, sky location, etc.), the full posterior for all parameters given both sources of data, $p(\vec{\kappa}, \vec{\xi} \mid D_{GW},D_{EM})$, can be factorized using Bayes' theorem as
\begin{align}
    p(\vec{\kappa}, \vec{\xi} | D_{GW},D_{EM}) &\propto p(\vec{\kappa}, \vec{\xi} \mid D_{EM})\, p(D_{GW} | \vec{\kappa}, \vec{\xi},D_{EM})\, , \nonumber \\
    &\propto p(\vec{\kappa})\, p(\vec{\xi} | D_{EM})\, \mathcal{L}(D_{GW} | \vec{\kappa}, \vec{\xi})\, ,
    \label{eq:posterior}
\end{align}
where $p(\vec{\kappa})$ is our prior for the birefringence parameters, $p(\vec{\xi} \mid D_{EM})$ represents our expectation for the binary properties conditioned on the EM data, and $\mathcal{L}(D_{GW}\mid\vec{\kappa},\vec{\xi})$ is the GW likelihood.
In deriving Eq.~\eqref{eq:posterior}, we have assumed that the EM data do not directly inform our knowledge of $\vec{\kappa}$, such that $p(\vec{\kappa}, \vec{\xi}\mid D_{EM}) = p(\vec{\kappa}\mid\vec{\xi}) \, p(\vec{\xi}\mid D_{EM})$, and that our prior on $\vec{\kappa}$ is independent from $\vec{\xi}$, such that $p(\vec{\kappa}\mid\vec{\xi}) = p(\vec{\xi})$; we have also used the fact that the GW data are generated independently from the EM data, so that $p(D_{GW}\mid\vec{\kappa},\vec{\xi}, D_{EM}) = p(D_{GW}\mid\vec{\kappa},\vec{\xi}) \equiv \mathcal{L}(D_{GW}\mid\vec{\kappa},\vec{\xi})$.
As usual, the corresponding (marginal) constraint on birefringence is obtained by marginalizing over nuisance parameters, i.e.,
\begin{equation}
    p(\vec{\kappa} \mid D_{GW}, D_{EM}) = \int d\vec{\xi}\, p(\vec{\kappa}, \vec{\xi} | D_{GM}, D_{EM})\, .
\end{equation}

As seen in Eq.~\eqref{eq:posterior}, the EM data may only enter our analysis through their implications for our knowledge of the binary properties $\vec{\xi}$.
Concretely, the EM data may constrain the source sky location, distance, inclination and orientation, so that
\begin{equation}
    p(\vec{\xi} \mid D_{EM}) = p(\vec{\theta})\, p({\rm RA}, {\rm Dec}, d_c, \iota, \psi \mid D_{EM})\, ,
\end{equation}
where $\vec{\theta}$ here represents all binary parameters not informed by the EM data (like the masses and spins), namely $\vec{\theta} \equiv \vec{\xi} \setminus \{{\rm RA}, {\rm Dec}, d_c, \iota, \psi\}$. In other words, $p(\theta)$ represents our prior on masses, spins, phase and time of arrival, and $p({\rm RA}, {\rm Dec}, d_c, \iota, \psi \mid D_{EM})$ represents the EM measurement of the extrinsic properties.
When we choose to use GW data alone without EM information, we simply replace this last factor by the prior $p({\rm RA}, {\rm Dec}, d_c, \iota, \psi \mid D_{EM}) \to p({\rm RA}, {\rm Dec}, d_c, \iota, \psi)$, which amounts to assuming the EM data do not constrain any binary parameters.

If no EM information is used, we choose the $\vec{\xi}$ priors to be the same as those used in \cite{Wong:2023lgb}. Based on the discussion in Sec.\ \ref{sec:EM}, when incorporating EM information we adjust the priors as follows. For the GW170817 event, when the EM sky location is included, we fix the values  $({\rm RA},{\rm Dec})= (13{:}09{:}48.085,-23{:}22{:}53.343)$ \cite{LIGOScientific:2017ync}, so that the prior is a delta function for those parameters. To factor in EM redshift/distance information, we use a Gaussian distribution in $d_L(z)$ with mean $42.9\,$Mpc and standard deviation 
of $3.2\,$Mpc \cite{LIGOScientific:2017zic}. For the inclination angle, we approximate the EM measurement as a Gaussian with mean $2.85\,$rad and standard deviation $0.03\,$rad \cite{Hotokezaka:2018dfi}. 
Finally, for the source orientation $\psi$, we use a Gaussian distribution with mean 3.14 rad and standard deviation of 0.09 rad based on our EM measurement from Sec.~\ref{sec:EM}.

In addition, for the priors on $\kappa_{A,V}$ we use the consistency bounds of Eqs.\ \eqref{kA_bound}--\eqref{kV_bound}. In the case of the GW170817 event, the maximum-likelihood distance is $d_c\approx 43\,$Mpc based on EM information, and the highest meaningfully detectable GW frequencies are $f\sim 1000$Hz. We thus assume a flat prior on $\kappa_A$ in the range $\kappa_{A}\in[-4,4]$.
For velocity birefringence, we assume a flat prior in the range  $\kappa_{V}\in[-164,164]$ rad, when using the best-fit redshift of the source from EM observations of $z=0.0096$ \cite{LIGOScientific:2017zic}.

Even though we measure them jointly, in the following we will show marginal posteriors for $\kappa_A$ and $\kappa_V$ separately. This is because we find these measurements to be effectively uncorrelated, and thus the results do not change when we consider them simultaneously.

The GW likelihood $ \mathcal{L}(D_{GW}|\vec{\xi},\vec{\kappa})$ is sampled through a matched-filtering exploration of the parameter space, given a waveform model, and assuming additive stationary Gaussian noise colored by a known power-spectral density (PSD). 
For the GW170817 event, we use the fast high-performance code \textsc{jim} \cite{Wong:2023lgb, Wong:2022xvh} (and follow their sampling choices), which performs Bayesian inference using automatically-differentiable waveform templates. With this code, we use the quadrupole-only \textsc{IMRPhenomD} template \cite{Husa:2015iqa, Khan:2015jqa}, which assumes that the individual neutron stars spins are aligned with their orbital angular momentum vectors, ignores tidal effects, and is available in a differentiable form through \textsc{ripple} \cite{Edwards:2023sak}. In reality, it is known that BNS mergers also exhibit tidal effects that modify the GW waveform, but we do not expect those effects to change our birefringence results since they have fundamentally different frequency evolution \cite{Dietrich:2017aum} and only affect the near-merger/merger time window, contrary to the birefringence effects that modify the entire waveform even during early inspiral.
In addition, the quadrupole-only assumption in \textsc{IMRPhenomD} considers only a $(2,2)$ angular harmonic contribution, which is expected to hold for most BNS systems since their mass ratio is expected to be close to unity, and they are expected to have no precession (due to small spin magnitudes and no eccentricity).

In addition to analyzing GW170817, we also perform forecasts on mock BNS events. In those cases, the waveform template used will be \textsc{PhenomDNRT} \cite{Husa:2015iqa, Khan:2015jqa, Dietrich:2017aum}, which is a quadrupole-only waveform that assumes aligned spins but does incorporate tidal effects. For simulated events we perform parameter estimation with the \texttt{GW Analysis Tools} code \cite{Perkins:2021mhb}, modified to include amplitude and velocity birefringence in the waveform. When incorporating EM information, we use the same approach as with the real data by choosing appropriate priors on $\vec{\xi}$, as discussed above, informed by the discussions in Sec.\ \ref{sec:EM}.

For each mock event, we assume that the EM observations were produced by current facilities, while we vary the GW detector network. We consider three different GW detector configurations. First, a 2nd-generation network consisting of LIGO-Hanford, LIGO-Livingston, Virgo, KAGRA, and LIGO-India, all presumed to be operating with LIGO A+ sensitivity \cite{Cahillane:2022pqm} (this configuration will be denoted `2G'). Then, we investigate the prospects for 3rd-generation (3G) detector sensitivities, such as Cosmic Explorer (CE). We consider a scenario that includes the five 2G detectors with A+ sensitivity and one CE detector at the Great Basin in Nevada (configuration denoted `2G + CE'). Finally, we consider a scenario in which there are two CE detectors in the United States, with one located in the Great Basin  \cite{Perkins:2021mhb} and one at the current LIGO-Hanford site (configuration denoted `CE').

\subsection{Results using the GW170817 event}\label{sec:170817}

\subsubsection{Amplitude birefringence}

\begin{figure}[h!]
\centering
\includegraphics[width = 0.40\textwidth]{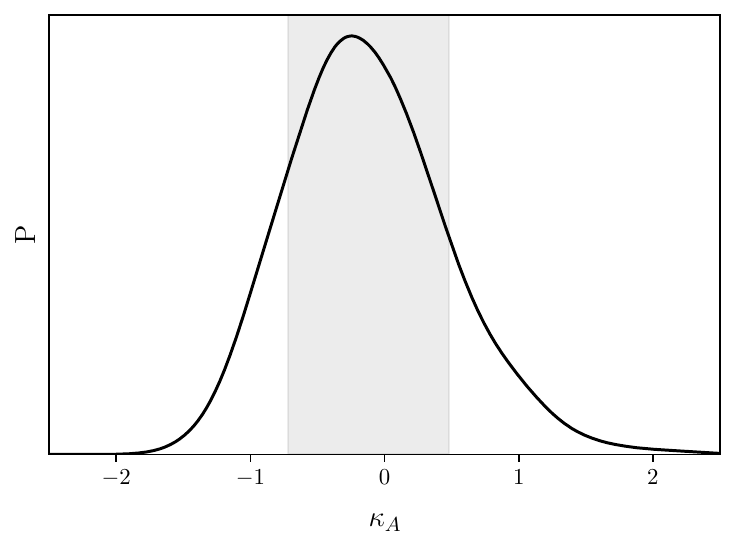}
 \caption{$\kappa_A$ posterior for GW170817, when using all EM information available. Grey band shows the $68\%$ credible interval. }
\label{fig:kappaA}
\end{figure}

Let us start by discussing amplitude birefringence. Figure\ \ref{fig:kappaA} shows the marginalized posterior on $\kappa_A$ when including all the EM information available (sky location, inclination, distance/redshift, orientation angle), which yields $\kappa_A=-0.12^{+0.60}_{-0.61}$ at 68\% credibility. The value of $\kappa_A=0$ lies within $1\sigma$, which implies consistency with GR.

Previous works have used instead the entire BBH population from the LVC to constrain amplitude birefringence. As a comparison, Ref.~\cite{Ng:2023jjt} used 71 BBHs to obtain $|\kappa_A|\lesssim 0.03$ with 68\%CL,\footnote{The relation between the parameter $\kappa_A$ defined here and the parameter $\kappa$ defined in \cite{Ng:2023jjt} is $\kappa_A = \kappa \times 1\,$Gpc.} which is about one order of magnitude better than the constraints obtained by the GW170817 event alone presented here. In fact, Table I of \cite{Ng:2023jjt} shows that several individual BBH events alone give tighter constraints on amplitude birefringence than the GW170817 event. This is because the polarization effects considered here accumulate over cosmological distances, such that BBH events with lower SNR at greater distances can be more constraining than higher SNR but nearby BNS events.  As an example, one of the BBH events that gives the best constraints on $\kappa_A$ is GW200129{\_}065458 \cite{LIGOScientific:2021djp}, which had an SNR of 26.8 (compared to GW170817 with SNR of 32.4) but occurred at $d_L\approx 900\, $Mpc (compared to $d_L\approx 43\,$Mpc for GW170817). 

\begin{figure}[h!]
\centering
\includegraphics[width = 0.40\textwidth]{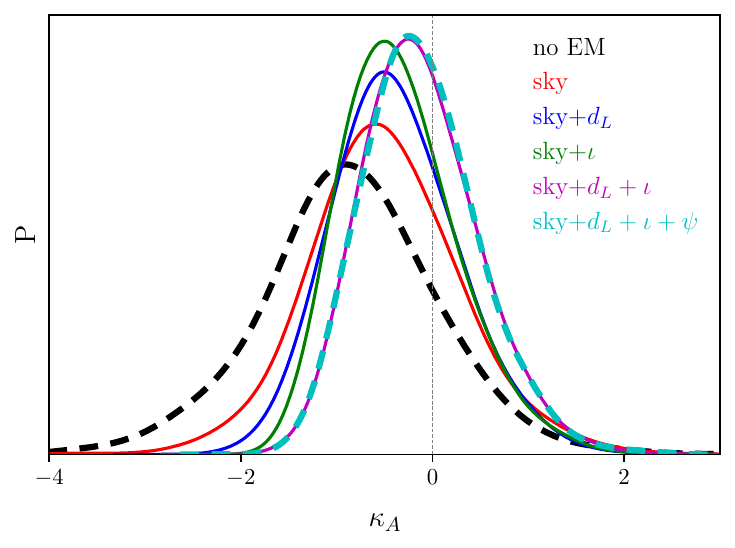}
 \caption{Normalized posterior distribution of $\kappa_A$ when varying the EM information input in the analysis. The dashed lines show the results when including no EM information (black) and all EM information available (cyan).}
 \label{fig:kappaA_posterior}
\end{figure}

Next, we analyze how different pieces of EM observations contribute to the $\kappa_A$ constraints. Figure\ \ref{fig:kappaA_posterior} shows the constraints on $\kappa_A$, comparing the posterior distributions in six cases:  when no EM info is used (black, dashed); when only the EM sky localization is used (red); when the sky and distance (redshift) EM information is used (blue);  when the sky and inclination information is used (green); when sky, distance and inclination EM information is used (magenta); and when all EM information is used (cyan, dashed), as in Fig.~\ref{fig:kappaA}. Using no EM information, we obtain $\kappa_A=-0.87^{+0.90}_{-0.88}$ at $68\%$ credibility, which is about $1.5\times$ broader than the constraint using all EM information ($\kappa_A=-0.12^{+0.60}_{-0.61}$) and further shifted away from GR.
From Fig.~\ref{fig:kappaA_posterior} we see that all individual pieces of EM information contribute to improve the $\kappa_A$ constraint, except for $\psi$.
This is expected because, for nearly face-on/off binaries dominated by the $\ell=|m|=2$ mode, $\psi$ only enters the waveform as an overall phase and is uncorrelated with amplitude birefringence. 
For this event, and assuming the sky position is known, the EM constraint on $\iota$ provides a marginally stronger leverage on $\kappa_A$ than the distance $d_L$, due to its tight precision and its effect on the amplitude of the signal.\footnote{Because $\iota$ and $d_L$ are strongly degenerate, constraints on $\iota$ from EM observations constrain $d_L$ from GW observations to a similar degree. For this event, the direct EM information on $\iota$ is more constraining than the direct EM information on distance/redshift, which is more loosely determined}.

Even though, overall, the improvement on the precision of $\kappa_A$ is not that considerable when including EM observations (a ${\sim}50\%$ improvement), we find that the EM data helps break mild parameter degeneracies and shift the best (median) estimate of $\kappa_A$ increasing posterior support for GR. For instance, this happens when including sky localization (red curve in Fig.\ \ref{fig:kappaA_posterior}), which is also correlated with distance, since the angles $(\theta,\phi)$ affect the amplitude of the observed GW signal and could thus be partially compensated by $\kappa_A$.

\begin{figure}[h!]
\centering
\includegraphics[width = 0.40\textwidth]{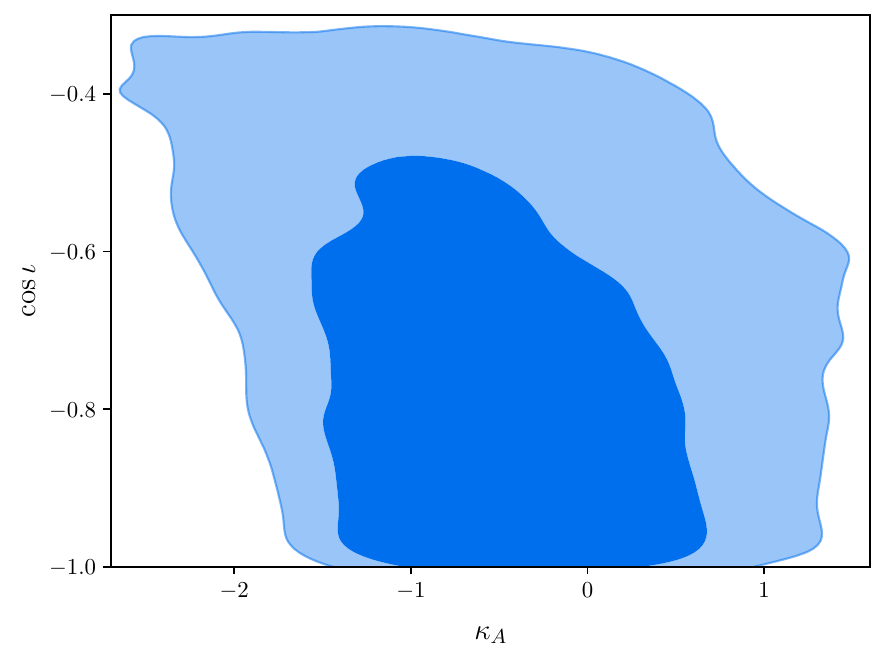}
 \caption{Joint $\kappa_A$  and $\cos \iota$ posteriors when using only sky EM information. The contours show 68\% and 95\% CL. No strong correlation is observed.}
\label{fig:kappaA_vs_iota_posterior}
\end{figure}

As previously mentioned, if amplitude birefringence does not exhibit any frequency dependence, then a full degeneracy exists between $\kappa_A$ and ($\iota$, $d_L$), which can be broken for individual GWs events that contain higher angular harmonics or by performing population-level inference such as in Ref.~\cite{Okounkova:2021xjv}. However, due to the frequency dependence of amplitude birefringence, we confirm that for BNS sources there is no strong degeneracy with $\iota$, as shown in Fig.~\ref{fig:kappaA_vs_iota_posterior} (nor with $d_L$). For this reason, when including EM information on $\iota$ or $d_L$ in Fig.\ \ref{fig:kappaA_posterior}, we only observe an increase in the precision of $\kappa_A$ but no shift on its central value (compare e.g.\ green and red curves).
Since EM-based constraints on the binary inclination are dependent on uncertain details of the jet modeling \cite{2021Univ....7..329L}, this lack of correlation implies that our ability to constrain amplitude birefringence will fortunately be not strongly susceptible to potential biases introduced by jet modeling uncertainties. 

\begin{figure}[h!]
\centering
\includegraphics[width = 0.40\textwidth]{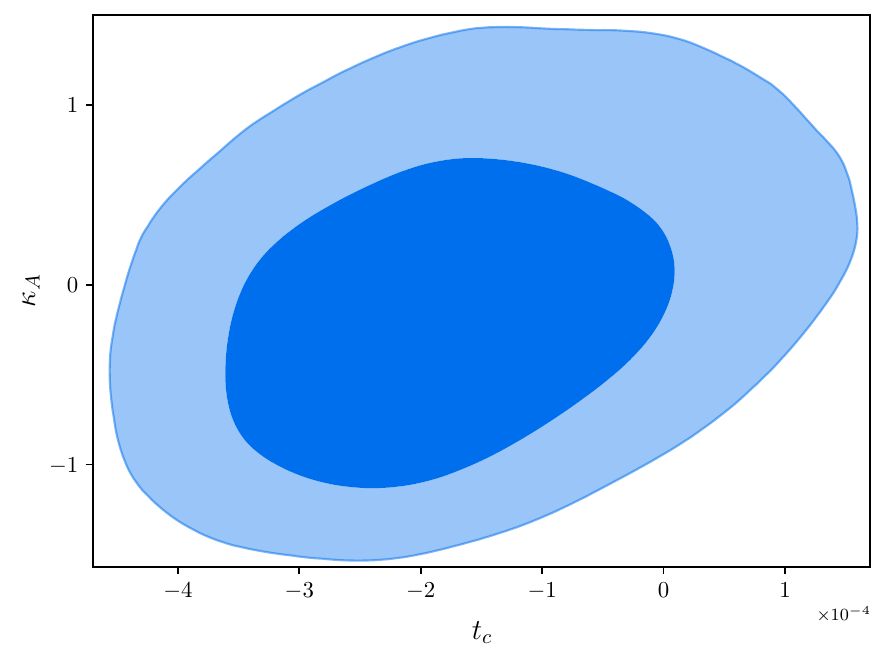}
 \caption{Joint $\kappa_A$  and $t_c$ posteriors when all EM information is used. The contours show 68\% and 95\% CL. 
 }
\label{fig:kappaA_vs_tc}
\end{figure}

Nevertheless, amplitude birefringence does exhibit some degeneracies with the coalescence phase $t_c$ for this event that was observed only by the LIGO detectors, as discussed in Sec.\ \ref{sec:GWbire}. We can see this in Fig.\ \ref{fig:kappaA_vs_tc}, where we find that a shift of order unity in the value of $\kappa_A$ can induce a shift of order $10^{-4}$sec in $t_c$ (see related discussion in Appendix \ref{sec:bilby}, where slight discrepancies with another numerical code were obtained for $t_c$ and $\kappa_A$, although the main conclusions remain the same). Nonetheless, these degeneracies can be broken with the use of multiple GW detectors with different orientations.

\subsubsection{Velocity birefringence}

Next, we discuss velocity birefringence. Since the models studied here introduce a frequency-independent velocity birefringence, this effect is completely degenerate with the orientation angle $\psi$ of the antenna pattern function, regardless of the GW source properties. That is, a change in $\psi$ can be compensated by a corresponding change in $\kappa_V$, which effectively means that we need to separately constrain the orientation angle at emission and at detection in order to measure $\kappa_V$ (cf., Eq.\ (\ref{Psi_degener2})); in other words, we need to obtain an independent EM measurement of the binary's orientation to compare to the orientation GW polarization ellipse measured on Earth..

\begin{figure}[h!]
\centering
\includegraphics[width = 0.40\textwidth]{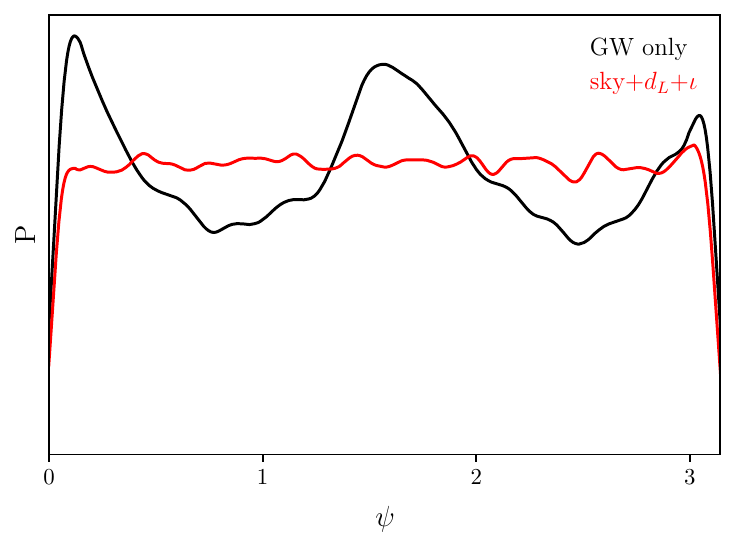}
 \caption{Posterior distribution for $\psi\in [0,\pi]$ rad in GR, using no EM information (black) and using sky, distance and inclination EM information (red). We see that the posterior distribution is uninformative. }
\label{fig:Psi_GR}
\end{figure}

Since the polarization orientation angle is not currently well constrained by GW detectors (especially for this almost circularly polarized signal), we have little information on the orientation of the detected GW polarization ellipse, and hence expect that $\kappa_V$ will also be similarly poorly constrained. Figure\ \ref{fig:Psi_GR} shows the marginalized posterior for $\psi$, which proves that, indeed, $\psi$ cannot be currently well measured, even when using multiple detectors. This is because the current LIGO detectors are nearly parallel, and although Virgo offers a different orientation, its detection of the GW170817 event occurred with much lower sensitivity (in fact, the GW170817 event had no SNR above the detectability threshold for Virgo \cite{LIGOScientific:2017vwq}); additionally, this source was nearly face-off, so. For this reason, we use the GW170817 event as a proof of principle to study constraints on $\kappa_V$, but we do not expect to find informative constraints, even when having very precise EM constraints on $\psi$.

\begin{figure}[h!]
\centering
\includegraphics[width = 0.40\textwidth]{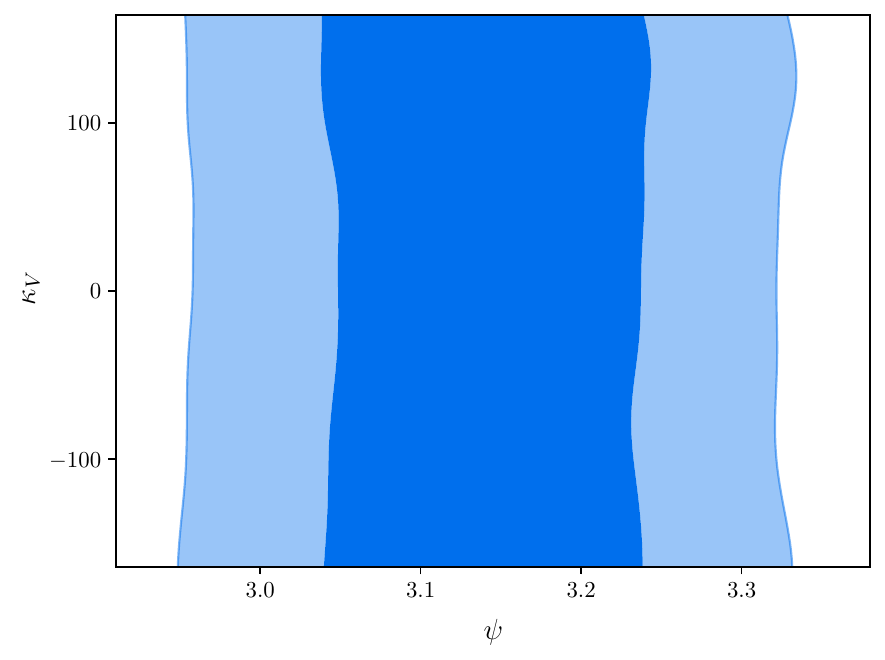}
 \caption{Joint $\kappa_V$ [rad]  and $\psi$ [rad] posterior distributions using all EM available information. The contours show the $68\%$ and $95\%$ CL. }
\label{fig:kappaV_PV}
\end{figure}

Figure \ref{fig:kappaV_PV} shows the joint posterior distributions of $\kappa_V$ and $\psi$ for the GW170817 event using all the available EM information (sky location, distance, inclination, and orientation angle). As expected, while the $\psi$ distribution\footnote{While it is customary for $\psi$ to be defined in the $[0,\pi]$ range as in Fig.\ \ref{fig:Psi_GR} due to its $\pi$ periodicity, for visual convenience we have extended the range of $\psi$ in Fig.\  \ref{fig:kappaV_PV}.} is determined by the EM information, $\kappa_V$ is completely unconstrained because these GW data alone does not provide information about the effective GW orientation angle (cf.\ Fig.\ \ref{fig:Psi_GR}). Note that the $\kappa_V$ range shown in Fig.\ \ref{fig:kappaV_PV} is such that $\delta\phi_V\in [-\pi,\pi]$ rad assuming $z=0.0096$. Because $\kappa_V $ is unconstrained, we find that the posteriors between $\psi$ and $\kappa_V$ are effectively independent of each other.

\begin{figure}[h!]
\centering
\includegraphics[width = 0.40\textwidth]{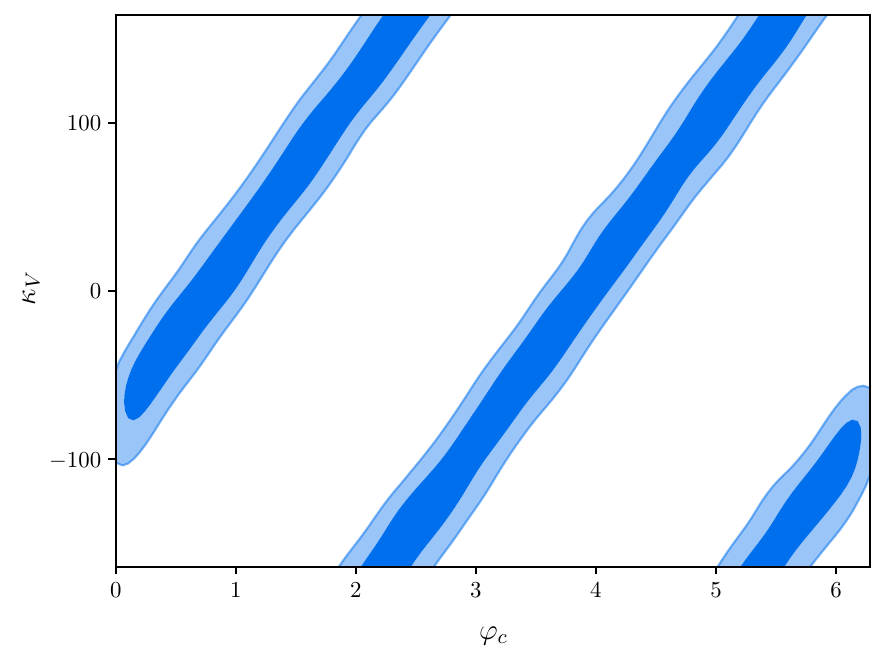}
 \caption{GW170817 $\kappa_V$  and $\varphi_c$ posterior distributions using all EM available information. The contours show the $68\%$ and $95\%$ CL. }
\label{fig:kappaVphic_PV}
\end{figure}

Furthermore, Fig.~\ref{fig:kappaVphic_PV} illustrates a degeneracy between $\kappa_V$ and $\varphi_c$, which arises from this event being viewed from a nearly face-off orientation, as discussed in Sec.\ \ref{sec:velbire}.  This joint posterior is given by multiple stripes, which arise because $\varphi_c$ exhibits a periodicity of $\pi$ due to this signal being dominated by the $(\ell=2, |m|=2)$ spherical angular harmonic (cf.\ Eq.\ (\ref{22Pol})).  For future events, detectors with improved sensitivity would help break this degeneracy by increasing the SNR and by potentially detecting additional subdominant angular harmonics.

\begin{figure*}[t]
\includegraphics[width=0.45\textwidth]{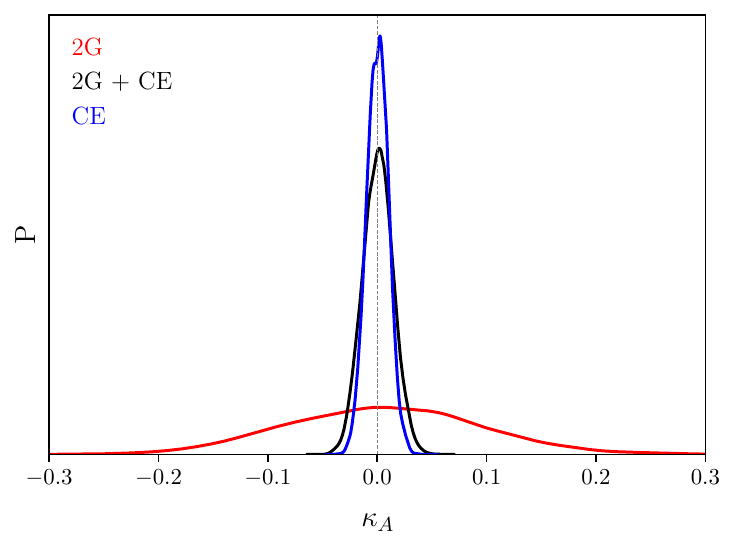}
\includegraphics[width=0.45\textwidth]{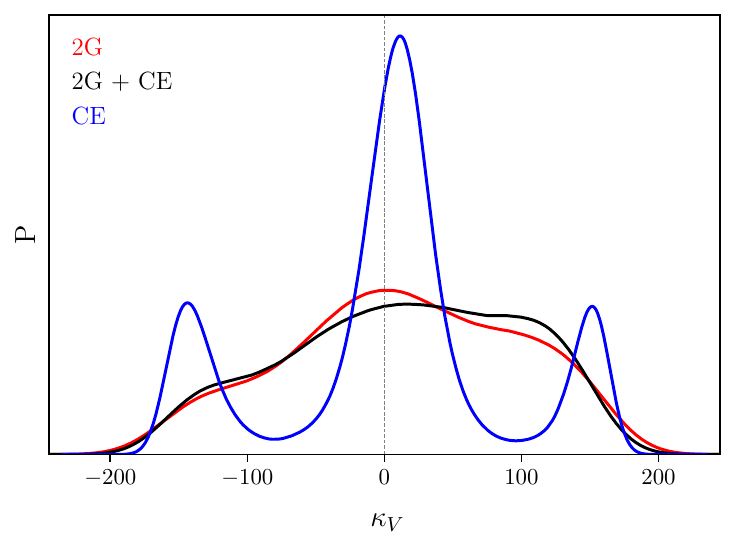}
    \caption{Posterior distribution on $\kappa_A$ (left) and $\kappa_V$ (right) for a mock BNS event with the same properties as GW170817 for three GW detector scenarios. Injected values shown in gray vertical lines.}
    \label{fig:kappaA_mock_170817}
\end{figure*}

\subsection{Future BNS events}\label{sec:mockBNS}
Having placed constraints on GW amplitude and velocity birefringence from GW170817, we now consider how future BNS events may improve these constraints.

\subsubsection{Future GW170817-like BNS events}

As previously discussed, the main limitation in probing velocity birefringence with GW170817 was the lack of GW detectors with different orientations, which left the polarization orientation completely unconstrained by the GW data. Motivated by this, in this section we mock a GW event with identical properties as GW170817 (same distance, inclination, masses, etc  \footnote{Specifically, the values we use are: masses $m_1= 1.46 M_\odot$, $m_2 = 1.27 M_\odot$, spins $\chi_1=\chi_2= 0.01$, distance $d_L = 42.9$ Mpc, orientation angle $\psi=3.14$ rad, inclination angle $\iota = 2.85$ rad, sky position $RA=3.45$ rad, and $Dec =-0.39$ rad, and coalescence phase $\varphi_c = 0.75$. The tidal information is contained within the parameter $\bar{\lambda}_s$ \cite{Yagi:2013bca}, which we take to be $\bar{\lambda}_s = 200$.}) and same EM information as priors.  We only change the GW detector network in order to explore how much the GW constraints will be improved in such a future scenario.

Figure \ref{fig:kappaA_mock_170817} shows the results for amplitude (left panel) and velocity birefringence (right panel), respectively, for each of the three detector scenarios. In the `2G' scenario (red), the BNS event would have a SNR of 193.3 and yield 68\% CL constraints of $\kappa_A = 5.4\times 10^{-3}{}^{+0.088}_{-0.091}$ and $\kappa_V = 14.0^{+97.2}_{-89.3}$ rad. Adding a Cosmic Explorer detector to obtain the `2G + CE' configuration (black) would yield an event with SNR of 2459.7 and constraints of $\kappa_A = 4.3 \times {10^{-4}} \pm 0.014$ and $\kappa_V = 15.2^{+98.7}_{-95.5}$ rad. Finally, considering a network of two Cosmic Explorers, the `CE' scenario (blue) yields an event with an SNR of 3454.1 and constraints of $\kappa_A = 2.6\times 10^{-4} \pm 0.010$ and $\kappa_V = 8.7 \pm 28.4$ rad, where we have quoted the $1\sigma$ width of the central feature. 

From these results we can extract a few conclusions. First, we observe that the maximum likelihood values of the marginalized posterior on $\kappa_A$ and $\kappa_V$ are both consistent with zero, which corresponds to the injected value, and hence there is no bias. The two-sided constraints on $\kappa_A$ then go from $\approx 0.18$ for the 2G configuration to $\approx 0.03$ for the 2G+CE configuration (a factor of 6 better), and to $\approx 0.02$ for the CE configuration (a further factor of 50\%). Due to the low distance to this event, even in the 2G scenario with SNR $\sim 190$, the constraint obtained on $\kappa_A$ is still weaker than the current one obtained by stacking BBH events \cite{Ng:2023jjt}. Nevertheless, the CE scenario increases the SNR by one order of magnitude, which translates also into a $\kappa_A$ constraint about one order of magnitude better.

\begin{figure}[b!]
\includegraphics[width=0.4\textwidth]{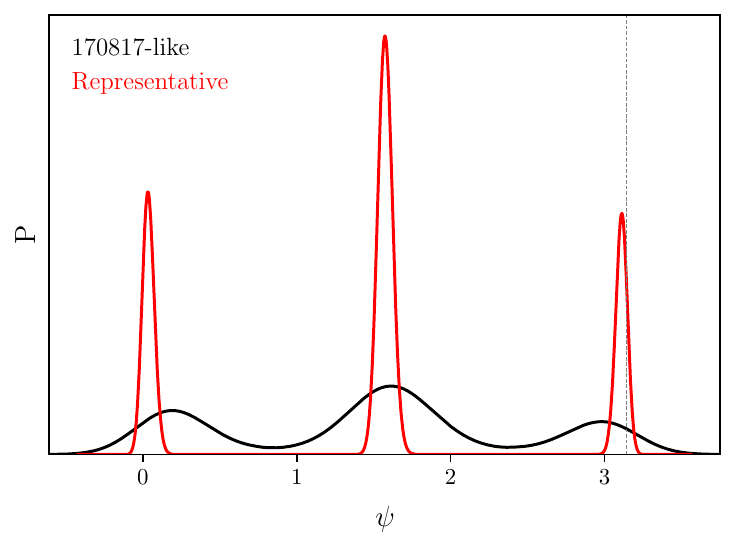}
\caption{Posterior distribution on $\psi$ [rad] for a GW170817-like event with the CE detector configuration in GR (no birefringence). Injected value is shown in gray vertical line.}
\label{fig:MockGRPsi}
\end{figure}

\begin{figure*}
\includegraphics[width=0.48\textwidth]{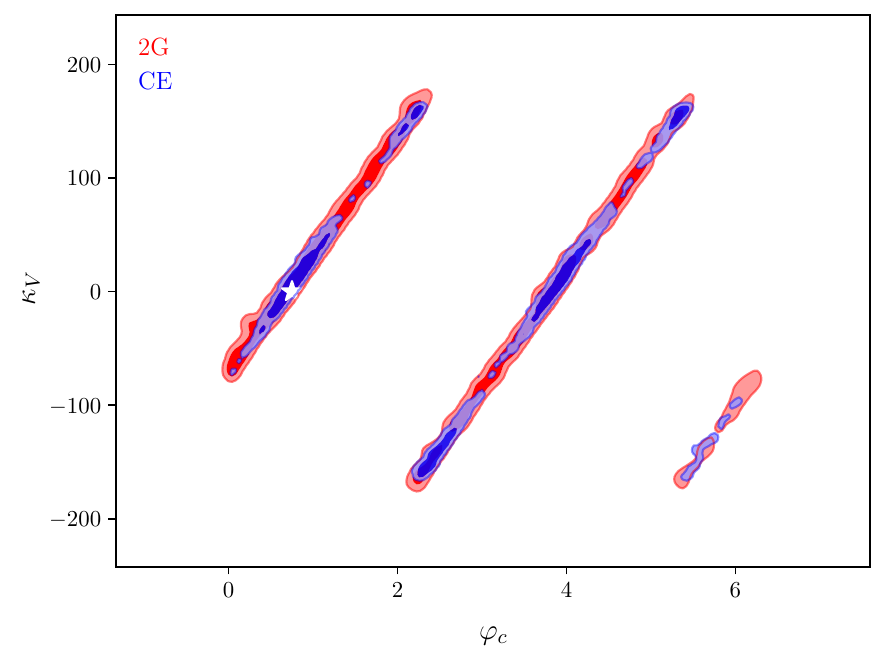}
\includegraphics[width=0.48\textwidth]{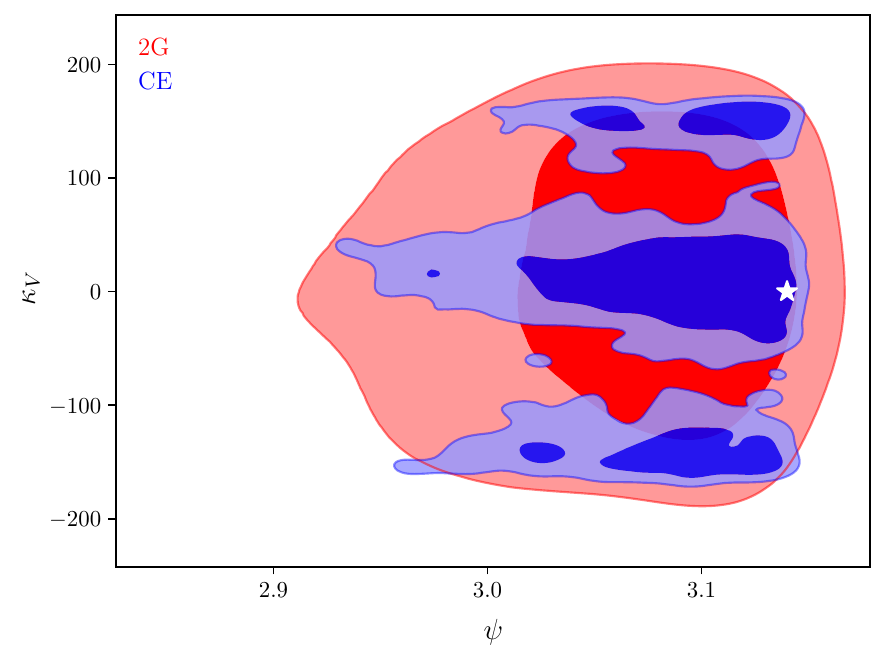}
    \caption{Joint $\kappa_V$ and $\varphi_c$ (left) and $\kappa_V$ and $\psi$ (right) posteriors for a mock GW170817-like event comparing the 2G and CE detector configuration. Injected values are indicated with a white star.}
    \label{fig:mock_phicAP}
\end{figure*}

For velocity birefringence, $\kappa_V$, we now see that the additional detectors with different orientations give meaningful constraints for a situation with high enough SNR in the CE scenario. Contrary to amplitude birefringence, the constraint on $\kappa_V$ now does not scale inversely proportional with SNR, which causes the improvement from the 2G to the CE scenario to be only a factor of $\sim$ three. This happens because one generally expects the strain precision to decrease linearly with SNR, but this depends on trigonometric functions of $\delta\phi_V$ instead of a linear function of $\delta\phi_V$, and the presence of additional degenerate angular parameters also degrade the $\kappa_V$ constraints. In the best-case scenario of CE, we obtain a precision on velocity birefringence equivalent to $|\delta\phi_V|< 0.7\,$rad at $68\%$ CL, for this event at $\sim40\,$Mpc.

Figure \ref{fig:MockGRPsi} shows the posterior on $\psi$ for this mock event (black line) in GR (assuming no birefringence but using the EM information for sky, inclination, and distance), in the CE scenario. We see that the $68\%$ CL uncertainty on $\psi$ is $\delta\psi \approx 0.27$ rad (estimation for each peak), which is weaker than the EM $\psi$ constraint (cf.\ Fig.\ \ref{fig:EMPsi2}). This aligns with our expectation that the precision on $\kappa_V$ is determined by the precision on $\psi$ (see Eq.\ \ref{Psi_degener2}), recalling that the degeneracy is such that $\psi$ is shifted by a factor of $\delta\phi_V/2 = \kappa_V \ln (1 + z)$. This means that for such highly aligned events, GW observations will be the limiting factor for performing tests of velocity birefringence. 
Notice that in the range of $\psi\in [0,\pi]$, the posterior from the GW observation is double peaked, which causes a similar distribution in $\kappa_V$ in Fig.\ \ref{fig:kappaA_mock_170817}. This multimodal behavior has been discussed in \cite{Roulet:2022kot}, where an approximate discrete symmetry of $\psi\rightarrow \psi+\pi/2$ is present for $(2,2)$-harmonic dominated waveforms, such as the one analyzed here.

Regarding parameter degeneracies, we now find that, due to the larger number of detectors, the degeneracy between $\kappa_A$ and $t_c$ is not present in any of the three detector scenarios. However,  
a degeneracy between $\kappa_V$ and $\varphi_c$ still remains. Figure \ref{fig:mock_phicAP} shows the joint $\kappa_V$ and $\varphi_c$ posteriors (left panel) and $\psi$ posteriors (right panel) in the 2G and CE detector scenarios. On the left panel, we can see that the degeneracy is still present even with more detectors in the 2G case, due to the low inclination of this event. Even with  the higher SNR from CE detectors, a strong degeneracy remains. This is the reason why $\kappa_V$ has support nearly throughout the entire range in Fig.\ \ref{fig:kappaA_mock_170817}. On the right panel, we can see that there is no visible degeneracy between $\kappa_V$ and $\psi$, even though it is expected due to Eq.\ (\ref{Psi_degener}). This happens because the uncertainty on $\kappa_V$ is too large.

We emphasize that 3G detectors will be sensitive to frequencies as low as a few Hz, which will make BNS signals detectable for hours, allowing the source to move across the sky due to Earth's rotation. This will introduce a time dependence in the antenna pattern function that will allow us to better constrain the binary's angular parameters. For instance, sky localization can improve by a factor of $2\times$ \cite{Zhao:2017cbb, Chan:2018csa, Baral:2023xst} due to Earth's motion.
Since the mock events analyzed here do not incorporate detector motion, the precision on $\psi$ and hence $\kappa_V$ estimated from GW data is conservative in the 2G+CE and CE scenarios.

Finally, we note that events at such a low redshift as GW170817 are not expected to happen very often, and thus even with 3G detectors, we expect to detect one merger of this type every 10 years \cite{Borhanian:2022czq}. For this reason, we next analyze the potential for constraining birefringence of a BNS event at a greater distance. 

\subsubsection{Future representative BNS events}
\begin{figure*}
    \centering 
    \includegraphics[width=0.45\textwidth]{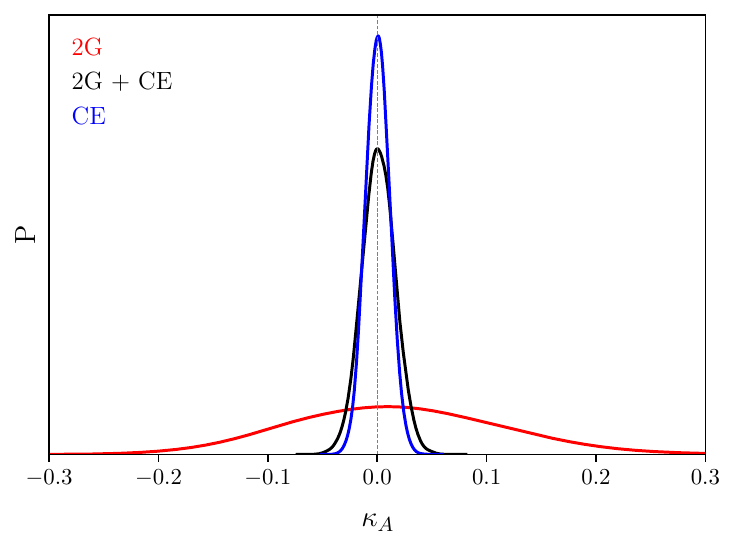}
    \includegraphics[width=0.45\textwidth]{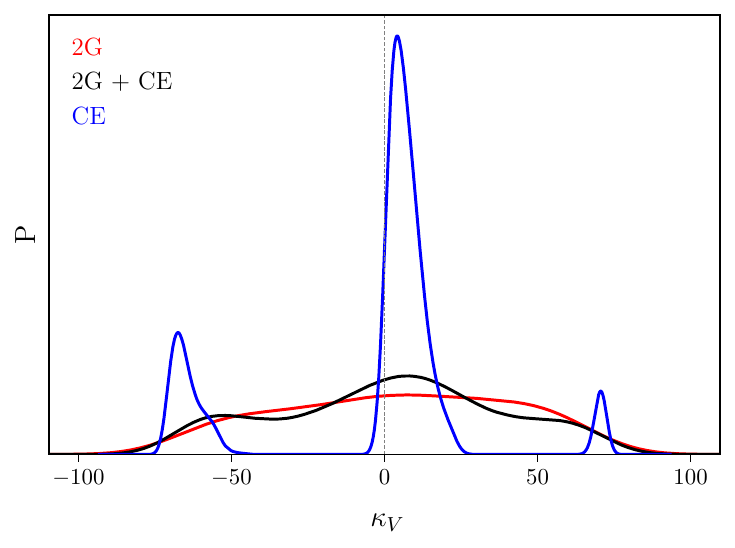}
    \caption{Posterior distribution on $\kappa_A$ (left) and $\kappa_V$ (right) for a mock BNS event inclined at $\theta_v = 30^\circ$ at 100Mpc for three GW detector scenarios.  As a reference, the entire range of $\kappa_V$ such that $\delta\phi_V\in [-\pi,\pi]$ is $\kappa_V\in [-72,+72]$. }
    \label{fig:kappaA_mock_iota30}
\end{figure*}

Next, we consider a more representative BNS, for which we assume an interstellar medium density of $n\sim 10^{-2}\textrm{cm}^{-3}$ more typical of those measured from short GRB afterglow modeling \cite{Fong+15} and expected from BNS population synthesis modeling \cite{Perna+22}, than the very low density inferred for GW170817. The interstellar medium density affects the afterglow brightness and evolution speed, and hence the precision on binary parameters obtained from EM observations (Sec.~\ref{sec:EM}).  

Whereas we will assume the same masses, spins, coalescence phase, and sky location as that of the GW170817 event, we now consider a binary with an angular momentum vector less aligned with the line of sight. In particular, we assume $\theta_v=30^\circ$ or, equivalently, $\iota=2.62\,$rad for a face-off case. This choice of inclination is favorable for GRB and afterglow detections but does not correspond to the most typical value of BNS inclinations expected for events within a given distance. In fact, for an isotropic inclination distribution, the median value is $\theta_v=60^\circ$ (corresponding to $|\cos\iota|=0.5$) and only $13\%$ of events would have $\theta_v\leq 30^\circ$. 

Having chosen an inclination value for the mock event, we will assume that EM observations provide the same fractional uncertainty in $\theta_v$ as that of the GW170817 event, so that $\iota=2.62 \pm 0.06$rad. However, we have checked that, due to the lack of correlations between $\iota$ and birefringence parameters, as well as the fact that $\iota$ is highly correlated with $d_L$ (and hence a $d_L/z$ observation provides information about $\iota$), the results do not change if one assumes no direct EM information on $\iota$ (and hence a flat prior in $\cos\iota$ for isotropic distributions).

In addition, we will assume the binary is at a distance of 100\, Mpc (or $z\approx 0.022$), where we may observe about one merger per year with 3G detectors \cite{Borhanian:2022czq} (which can then be stacked to improve further birefringence constraints) and still get good radio observations with current facilities. We will assume a redshift uncertainty of $\sigma_z\sim 10^{-3}(1+z)$ as expected from spectroscopic observations from nominal Euclid/DESI requirements \cite{2011arXiv1110.3193L, DESI:2016fyo} \footnote{The peculiar velocity correction is also expected to contribute with a similar uncertainty since  typical velocities are $v/c\sim 10^{-3}$.}.
Based on Fig.\ \ref{fig:theta_n} and the parameter choices previously discussed for this binary, we will then assume this binary to have associated EM orientation information with precision $\sigma(\psi) = 0.2$ rad and mean $\psi = 3.14$ rad. In addition, we will assume that EM observations will provide us with precise enough sky location so that effectively no uncertainty in these parameters is incorporated.  

The results for this mock event are shown in Fig.~\ref{fig:kappaA_mock_iota30} for amplitude and velocity birefringence. For the 2G scenario (red), this event will have SNR $= 75.2$ and would yield $\kappa_A=0.017\pm 0.098$ and $\kappa_V= 3.8^{+46.3}_{-47.5}$ rad at  68\% CL. For 2G+CE (black), we find an SNR of 961.3 and birefringence constraints of $\kappa_A={1.2\times 10^{-3}} \pm 0.015$ and $\kappa_V= 1.4^{+41.2}_{-46.6}$ rad. While for CE (blue), we find an SNR of 1345.0 and constraints of $\kappa_A={3.5\times 10^{-4}}\pm 0.011$ and $\kappa_V=6.8\pm 5.4$ rad.
We see that since this event is at a larger distance than the GW170817-like configuration, it gives comparable constraints for $\kappa_A$, and the larger inclination considerably improves the constraints for $\kappa_V$, for the same detectors, despite the lower SNR. 

For velocity birefringence, we obtain constraints on $\kappa_V$ such that $|\delta\phi_V|<0.24$ rad at $68\%$CL for CE, for this event at 100 Mpc. In this case, the precision is limited by the EM measurement of $\psi$. This can be seen from Fig.\ \ref{fig:MockGRPsi}, which shows the posterior on $\psi$ for this mock event (red line) in GR (assuming no birefringence but using the EM information for sky, inclination, and distance), in the CE scenario, which reaches a precision of about $0.04$ rad. This finding motivates considering the capabilities of next-generation radio telescopes in order to achieve further improvements on velocity birefringence constraints. In particular, the reduced beam size and much greater sensitivity of the planned ngVLA would greatly reduce the statistical error but observations would be dominated by systematic uncertainties, which can be reduced by a factor of a few.

\begin{figure*}
    \centering 
    \includegraphics[width=0.48\textwidth]{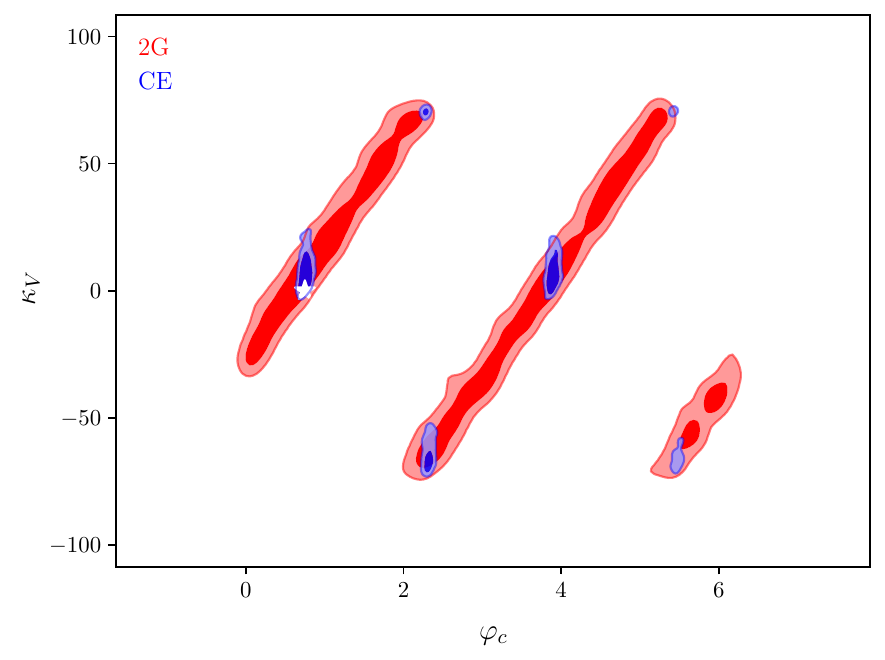}
   \includegraphics[width=0.48\textwidth]{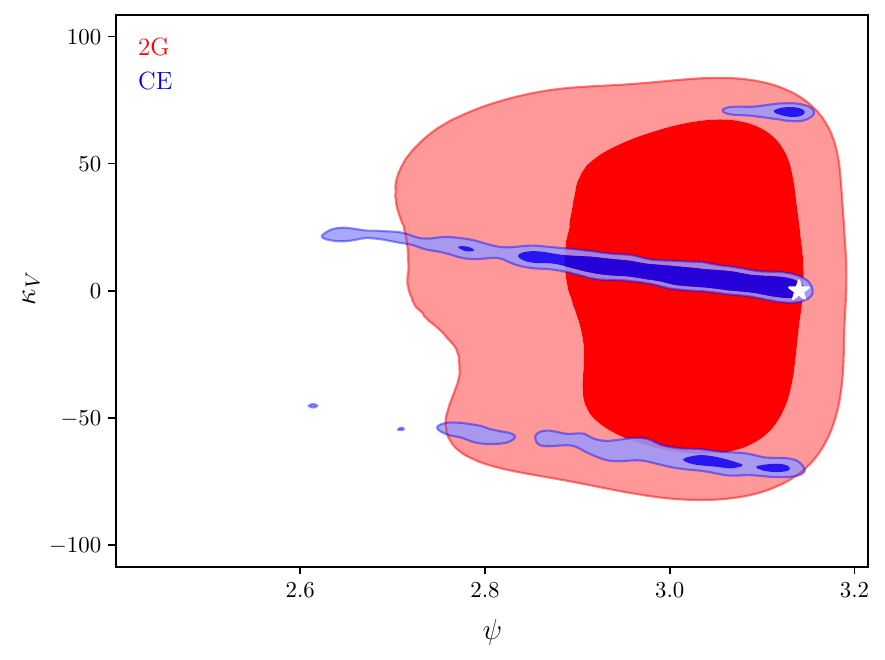}
    \caption{Joint $\kappa_V$ and $\varphi_C$ (left) and $\kappa_V$ and $\psi$ (right) posteriors for a mock BNS event inclined at $\theta_v = 30^\circ$ at 100Mpc comparing the 2G and CE detector configurations. Injected values are indicated with a white star.}
    \label{fig:kappaV_phi_c_mock_iota30}
\end{figure*}

Finally, Fig.\ \ref{fig:kappaV_phi_c_mock_iota30} shows the joint posteriors for $\kappa_V$ and $\varphi_c$ and $\psi$, for this 100 Mpc mock event (including all EM priors). On the left panel, we see that the larger inclination angle of this event leads to a smaller degeneracy between $\kappa_V$ and $\varphi_c$, as expected from the discussion in Sec.\ \ref{sec:velbire}. Indeed, for the CE configuration we find no degeneracy. This means that with these type of BNS events, velocity birefringence tests of GR will be less prone to biases. On the right panel, we now see an explicit degeneracy between $\kappa_V$ and $\psi$ for the CE configuration, given the better angular precision of the $\kappa_V$ measurement.

We summarize in Table \ref{table:summary} the constraints on the amplitude and velocity parameters, for GW170817 as well as the two BNS mock events when observed by the three different GW detector configurations considered in this paper.

\begin{table}[h!]
\caption{\label{table:summary}Summary of the constraints on the amplitude ($\kappa_A$) and velocity ($\kappa_V$) parameters for GW170817 and two BNS mock events across different GW detector configurations: 2G, 2G+CE, and CE.}
\begin{ruledtabular}
\begin{tabular}{lll}
\textbf{Event} & $\mathbf{\kappa_A}$ & $\mathbf{\kappa_V}$ \textbf{[rad]} \\
\hline\\[-5pt]
GW170817 & $-0.12^{+0.60}_{-0.61}$ & Unconstrained \\[4pt]
GW170817-like: 2G & $\sigma(\kappa_A)= 0.089$ & Unconstrained \\
GW170817-like: 2G+CE & $\sigma(\kappa_A)= 0.015$ & Unconstrained \\
GW170817-like: CE & $\sigma(\kappa_A)= 0.010$ & $\sigma(\kappa_V)= 28.4$ \\[4pt]
Representative BNS: 2G & $\sigma(\kappa_A)= 0.096$ & Unconstrained \\
Representative BNS: 2G+CE & $\sigma(\kappa_A)= 0.016$ & Unconstrained \\
Representative BNS: CE & $\sigma(\kappa_A)= 0.011$ & $\sigma(\kappa_V)= 5.4$ \\
\end{tabular}
\end{ruledtabular}
\end{table}

\section{Birefringence in Modified Gravity Theories}\label{sec:theory}
The constraints obtained on $\kappa_{A,V}$ in the previous section can be now translated to constraints on fundamental parity-breaking gravity theories, that introduce cosmological birefringence. Let us begin with the theory of dynamical CS gravity, which is characterized by a parity-violating coupling constant $\alpha_{CS}$, such that the Lagrangian density is (see \cite{Alexander:2009tp} for a review):
\begin{equation}
    \mathcal{L}[g,\theta]=\kappa R +\frac{\alpha_{\rm CS}}{4}\theta \; {}^{*}R R-\frac{1}{2}\nabla^\mu\theta \nabla_\mu\theta
\end{equation}
where $\kappa^{-1}=16\pi G$ using $c=1$, and $\theta$ is a dynamical scalar field that interacts with gravity via the dual Riemann term  ${}^{*}RR\equiv {}^{*}R_{\mu\nu\alpha\beta}R^{\nu\mu\alpha\beta}$ that breaks parity symmetry. In the literature, it is customary to use geometric units where $G=1$, in which case $\alpha_{\rm CS}$ has units of (Length)${}^{2}$. Previous studies on dynamical CS gravity have obtained upper bounds of $\alpha_{\rm CS}^{1/2}<\mathcal{O}(10^8)$ km based on Solar System (SS) observations and $\alpha_{\rm CS}^{1/2}<\mathcal{O}(10)$ km based on combined advanced LIGO and NICER observations, both of which assume no background or ``cosmological'' scalar field \cite{Yunes:2008ua, Ali-Haimoud:2011zme, Nakamura:2018yaw}; Ref.~\cite{Ng:2023jjt} constrained $\alpha_{\rm CS}^{1/2} \lesssim 40\, {\rm km}$ at 68\% credibility with LIGO-Virgo BBH observations. Other studies on CS gravity, that included a cosmological scalar field but ignored the inhomogeneous solution, placed constraints with SS observations on a combined length scale $\ell_{\rm CS}= \alpha_{\rm CS} \dot{\theta}/\kappa < {\cal{O}}(10^3)$ km \cite{Alexander:2007vt,Smith:2007jm}.

Modified gravity models may also induce distortions in the emission of GWs and we have ignored those in this paper. Currently, no constraints on dynamical CS gravity can be placed with such distortions alone because they are degenerate with spin effects~\cite{Nair:2019iur,Perkins:2021mhb, Wang:2021jfc}. 
Future observations of spin-precessing binary black hole inspirals, however, may be able to break these degeneracies and lead to constraints that are comparable to the combined advanced LIGO and NICER's one~\cite{Alexander:2017jmt,Loutrel:2022tbk}.

CS theory induces amplitude birefringence and hence  our $\kappa_A$ constraint from GW170817  translates to $\ell_{\rm CS}\approx \kappa_A \times 10^{3}{\rm km}<\mathcal{O}(10^3)$ km, which is similar to the SS bound obtained in \cite{Alexander:2007vt,Smith:2007jm}, when ignoring the inhomogeneous scalar field behavior. In order to make a comparison to the SS and NICER constraints on $\alpha^{1/2}$, some assumption about the scalar field cosmological evolution must be made. 
If we assume that the scalar field kinetic energy is cosmologically relevant, we estimate $\dot{\theta}\sim H_0/c$, and obtain $\alpha_{\rm CS}^{1/2}\sim \sqrt{c\ell_{\rm CS}/H_0}< \mathcal{O}(10^{13})$ km in geometric units. Note that if the scalar field was cosmologically irrelevant, then this constraint on $\alpha_{\rm CS}$ would be even weaker.

We see that this result is much weaker than current constraints, even if it were to increase by two orders of magnitude with another individual BNS observation with ground-based 3G GW detectors. Indeed, in the previous section we found that a single future BNS event will improve the constraint on $\ell_{CS}$ by two orders of magnitude, compared to that of GW170817, which is comparable to the constraints on $\ell_{CS}$ projected in \cite{Yunes:2010yf} by using BNS and their coincident $\gamma$-ray bursts. 
That the BNS constraint on CS gravity is weaker than other current constraints is not surprising given that this theory introduces modifications that increase with the spacetime curvature, and hence observations associated with stars or direct properties of compact objects are expected to be more sensitive than cosmological ones. Some modified gravity theories can introduce only low-curvature modifications if they are equipped with screening mechanisms  (see e.g.\ reviews \cite{Joyce:2014kja, Jain:2010ka}), but to date no such parity-violating theory has been studied in the literature.

Next, the constraint on $\kappa_V$ can be translated to Symmetric Teleparallel gravity \cite{Conroy:2019ibo}. This theory uses the Palatini approach, where the metric $g$ and the connection $\Gamma$ are independent, and its Lagrangian density is given by:
\begin{align}
    \mathcal{L}[g, \Gamma, \theta] = \frac{\kappa}{2}&\left[ -\frac{1}{2}Q_{\mu\alpha\beta}Q^{\mu\alpha\beta}+ Q_{\mu\alpha\beta}Q^{\alpha \mu \beta}+\frac{1}{2}Q_{\mu}Q^{\mu}\right. \nonumber\\
    &\left. -Q_{\mu}\tilde{Q}^{\mu}\right] +\mathcal{L}_{PV}[g, \Gamma, \theta],
\end{align}
where $Q$ describes the non-metricity tensor $Q_{\mu\alpha\beta}\equiv \nabla_{\mu}g_{\alpha\beta}$ (that vanishes in GR) with contractions $Q_{\mu}\equiv g^{\alpha\beta}Q_{\mu\alpha\beta}$ and $\tilde{Q}_{\mu}\equiv g^{\alpha\beta}Q_{\alpha\beta\mu}$. Here, the covariant derivatives are taken with respect to a general connection $\Gamma$. Similarly to CS gravity, this theory can contain an additional scalar field with parity-violating interactions to gravity in $\mathcal{L}_{PV}$, given by:
\begin{align}
    \mathcal{L}_{PV} = & -\frac{1}{2}\nabla^\mu\theta \nabla_\mu\theta + \kappa \alpha_1\nabla_\mu \theta \nabla^\nu\theta \; {}^{*}Q^{\mu\delta}{}_{\lambda} Q_{\nu\delta}{}^{\lambda}\nonumber\\
    + &  \kappa\alpha_2\nabla_\mu \theta \nabla^\mu\theta \; {}^{*}Q^{\nu\delta}{}_{\lambda} Q_{\nu\delta}{}^{\lambda},
\end{align}
where we have introduced the dual non-metricity tensor ${}^{*}Q^{\mu\delta}{}_{\lambda}\equiv \epsilon^{\alpha\beta\mu\delta}Q_{\alpha\beta\lambda}$, and the arbitrary  coupling constants $\alpha_{1,2}$. This theory induces velocity birefringence, such that \cite{Jenks:2023pmk} $\kappa_V \approx  (2\alpha_1+\alpha_2)\dot{\theta}_0^2$, where we have assumed that $\theta$ evolves slowly and hence have approximated its derivative $\dot{\theta}$ to the value today $\dot{\theta}_0$. For future BNS events like GW170817, we will then be able to impose constraints on the coupling constants of this theory of the order $|(2\alpha_1+\alpha_2)\dot{\theta}_0^2|<\mathcal{O}(1)$ rad. To our knowledge, no observational constraint on this theory has been imposed to date.

\section{Conclusions}\label{sec:conclusions}

In this paper we have tested cosmological birefringence on GW signals from binary neutron stars. In particular, we probed two polarization phenomena: amplitude and velocity birefringence, which introduce an amplitude and phase difference between left and right-handed circular polarizations, and are characterized by the parameters $\kappa_A$ and $\kappa_V$, respectively. While amplitude birefringence can be (and has been) tested with other binary sources without EM counterparts, we show that velocity birefringence can only be tested with spatially-resolved (i.e., jet-like) EM counterparts that allow for a detailed characterization of the binary's angular momentum orientation since that determines the phase of the GW polarization at the moment of emission. We quantified the birefringence constraints for the event GW170817 as well as two BNS mock events. A summary of the results is provided in Table \ref{table:summary}. 

For amplitude birefringence, we found that GW170817 gives a constraint $\kappa_A=-0.12^{+0.60}_{-0.61}$, which has the null value $\kappa_A=0$ within $1\sigma$ in agreement with GR. For this event, incorporating EM information improved the precision on $\kappa_A$ by a factor of 1.5.
Nevertheless, this constraint is roughly 10 times weaker than previous constraints obtained from stacking BBHs observations \cite{Ng:2023jjt}. This arises because cosmological birefringence grows with cosmological distance, and thus sources that are farther away are generally more favorable for testing this phenomenon. In addition, by mocking future BNS events, we found that the precision on $\kappa_A$ scales inversely proportional to the SNR, and hence $\kappa_A$ constraints are expected to improve by as much as 60 times with a single BNS observed with 3G GW detectors. 

For velocity birefringence, we obtained no informative constraint on $\kappa_V$ for the GW170817 event. This was due to the fact that, even though the EM radio observations from the afterglow places tight constraints on the binary's angular momentum orientation, it was not possible to test whether the polarization phase changed from emission to detection since the GW data did not provide any mean meaningful polarization constraint.
We then performed mock BNS events detectable with a network of five 2G GW detectors (LIGO-Hanford, LIGO-Livingston, Virgo, KAGRA, and LIGO-India) as well as 3G Cosmic Explorer detectors. We find that, due to angular parameter degeneracies and the low inclination of BNS events with observable jets, it seems to be only possible to constrain $\kappa_V$ with 3G GW detectors, which can reach a precision of $\sigma(\kappa_V)\sim 5$rad. This precision can be reached for events with inclinations comparable or larger than $30^\circ$, in which case we find the main limitation to be the EM precision on the observed binary's orientation angle. In the future, it will be useful to consider the capabilities of next-generation radio telescopes in order to achieve further improvements on velocity birefringence constraints.

We emphasize that this parameter $\kappa_V$ cannot be constrained with BBHs as it crucially requires from an EM counterpart to inform the polarization properties at emission. 
While this paper focused on BNS mergers, other multi-messenger sources such as the so-called verification binaries \cite{2018MNRAS.480..302K, Finch:2022prg}, corresponding to galactic double white dwarf binaries detectable with LISA \cite{LISA:2017pwj}, could in principle be used for similar purposes provided a measurement of the binary's angular momentum inclination and orientation can be made through EM observations. Nonetheless, we expect that such binaries will generally provide weaker constraints than BNS systems because the birefringence effect analyzed in this paper is of cosmological origin and thus grows with distance to the source, making extragalactic sources far more promising than galactic ones (as was the case described above for BBH vs BNS).  
In the case of future space-based GW detectors, such as LISA, the binary events are expected to have very high redshifts, but the detected frequency is in the mHZ band and amplitude birefringence grows linearly with frequency. A detailed analysis will be done in the future to compare ground and space-based detectors.

BH-NS binary systems with EM counterparts provide another possible multi-messenger source for constraining birefringence. For velocity birefringence tests, such sources are beneficial insofar that their unequal mass ratios may allow for the detection of higher angular harmonics other than $(22)$, which would allow to break degeneracies between the velocity birefringence parameter $\kappa_V$ and the coalescence phase $\varphi_c$. Nevertheless, similarly to BNS, these tests require jet-like emissions from highly face-on/off binaries, and only a small fraction  of all detectable BH-NS are predicted to generate such emission \cite{2021ApJ...923L...2F, Sarin:2022cmu, Biscoveanu:2023eyp}.

\acknowledgments
M.\ L.\ was supported by the Innovative Theory Cosmology fellowship at Columbia University. The work of L.\ J.\ is supported by the Kavli Institute for Cosmological Physics at the University of Chicago through an endowment from the Kavli Foundation and its founder Fred Kavli. K.\ H.\ was supported by JST FOREST Program (JPMJFR2136) and the JSPS Grant-in-Aid for Scientific Research (20H05639, 20H00158, 23H01169, 20K14513, 23H04900). B.\ D.\ M.\ is supported in part by the National Science Foundation (grant \# AST-2002577). N.Y.~was supported by the Simons Foundation through Award No. 896696 and the National Science Foundation (NSF) Grant No. PHY-2207650. The Flatiron Institute is supported by the Simons Foundation.  
This work was performed under the auspices of the U.S. Department of Energy by Lawrence Livermore National Laboratory under Contract DE-AC52-07NA27344.
The document number is \REVIEWRELEASENUMBER{}. 
This material is based upon work supported by NSF's LIGO Laboratory which is a major facility fully funded by the National Science Foundation.

\appendix
\section{Code Comparison}\label{sec:bilby}
\begin{figure}[h!]
\includegraphics[width=0.4\textwidth]{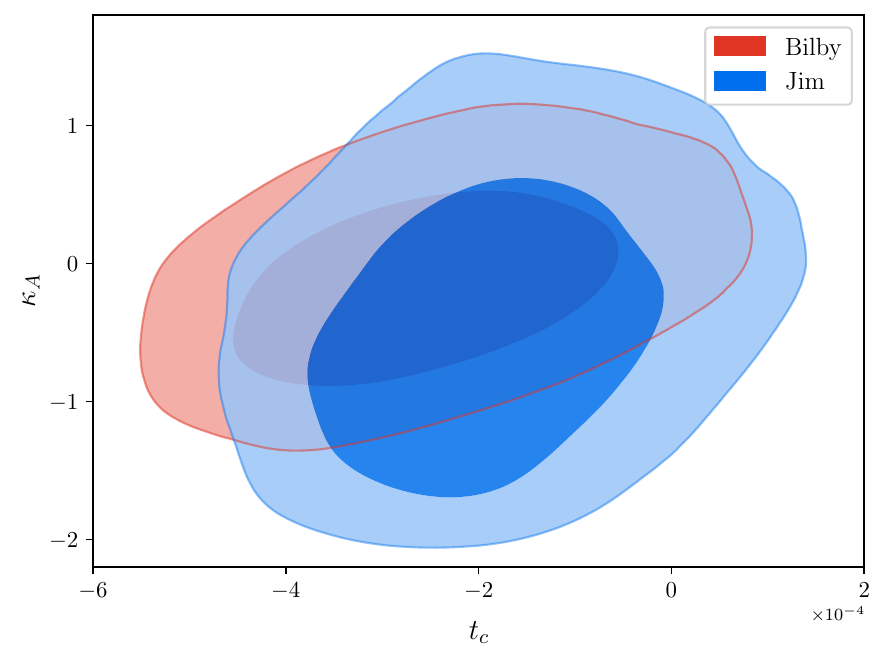}
    \caption{Joint $\kappa_A$ and $t_c$ [sec] posteriors for GW170817 obtained with Bilby and Jim, when fixing the sky localization of the source.}
    \label{fig:bilby}
\end{figure}

In order to test the robustness of the data analysis for GW170817, we also perform code tests for amplitude birefringence. We use the Bilby code \cite{Ashton:2018jfp} modified in \cite{Ng:2023jjt}, and compare to the results shown in Sec.\ \ref{sec:170817} using Jim. Using the waveform \texttt{IMRPhenomD}, we find all waveform parameters to be consistent, except for differences in $\kappa_A$ and $t_c$, as shown in Fig.\ \ref{fig:bilby}. 

Due to the difference in the level of degeneracy between these parameters in addition to some shifts, Bilby yields $\kappa_A=-0.14^{+0.47}_{-0.46}$ and $t_c=(-2.5\pm 1.3) \times 10^{-4}$ sec, whereas Jim yields $\kappa_A=-0.44^{+0.74}_{-0.75}$ and $t_c=(-1.9^{+1.1}_{-1.2})\times 10^{-4}$ sec at 68\% CL. After trying different code configurations, we have not obtained concluding evidence for what might be causing this difference. Nevertheless, both $\kappa_A$ values are within $1\sigma$ from each other, and thus the main results of this paper are not affected by this slight discrepancy. However, we highlight this issue here since this will require further investigation if these codes are wished to be used for future GW events with high SNR.

We emphasize that the agreement between Bilby and Jim for GR has been discussed in \cite{Wong:2023lgb}, where they were found to be in agreement. Nevertheless, that work did not discuss the results on $t_c$ but their data is publicly available and we find that a difference of  $0.6\times 10^{-4}$sec is also present in the mean of this parameter in GR, which is the same as to what we have obtained when including birefringence. In particular, in GR, Bilby yields $t_c=(4.5^{+4.5}_{-4.6})\times 10^{-4}$ sec and Jim yields $t_c=(5.1^{+4.6}_{-4.7})\times 10^{-4}$ sec for GW170817 using no EM information.
This suggests there might be a subtle difference between the codes, which is likely to be unrelated to the birefringence modifications we have made but it does impact the result for $\kappa_A$ due to its degeneracy with $t_c$.

\bibliography{Refs}

 \end{document}